\documentclass[preprint,5p]{elsarticle}
\usepackage{tikz}
\usetikzlibrary{shapes}
\usepackage{rotating}
\usepackage{subcaption} 
\usepackage{import}
\usepackage{textcomp}
\usepackage{units}
\usepackage{pgfkeys}
\usepackage{amsmath}
\usepackage{multirow}
\usepackage[hidelinks]{hyperref}

\newcommand{\executeiffilenewer}[3]{%
\ifnum\pdfstrcmp{\pdffilemoddate{#1}}%
{\pdffilemoddate{#2}}>0%
{\immediate\write18{#3}}\fi%
}

\usepackage{amssymb}

\usepackage{epstopdf}

\definecolor{cream}{RGB}{222,217,201}

\definecolor{clearblue}{RGB}{192,211,237}
\definecolor{magenta}{RGB}{72,204,217}
\definecolor{orange}{RGB}{253,141,60}
\definecolor{green}{RGB}{174,230,162}
\definecolor{yellow}{RGB}{227,227,165}
\definecolor{blue}{RGB}{78,148,196}
\definecolor{grey}{RGB}{213,213,213}
\definecolor{darkgrey}{RGB}{159,159,159}
\definecolor{darkblue}{RGB}{49,136,189}
\definecolor{lightblue}{RGB}{158,202,225}
\definecolor{middleblue}{RGB}{107,174,214}
\definecolor{darkorange}{RGB}{230,85,13}
\definecolor{lightorange}{RGB}{253,141,60}
\definecolor{middleorange}{RGB}{253,174,107}

\begin{document}
\begin{frontmatter}
  \title{On the validity of the Arrhenius picture in two-dimensional submonolayer growth}
  \author[1,2]{Joseba Alberdi-Rodriguez\corref{cor1}}
  \author[4]{Shree Ram Acharya}
  \author[1,4]{Talat S. Rahman}
  \author[2,3] {Andres Arnau}
  \author[1,2,3]{Miguel Angel Gos\'alvez}
  
  \cortext[cor1]{Corresponding author: joseba.alberdi@ehu.eus}
  
  \address[1]{Donostia International Physics Center (DIPC), Manuel Lardizabal
    4, 20018 Donostia-San Sebasti\'an, Spain}
  
  \address[2]{Departamento de F\'isica de Materiales, University of the
    Basque Country UPV/EHU, Manuel Lardizabal 3, 20018 Donostia-San
    Sebasti\'an, Spain}
  
  \address[3]{Centro de F\'isica de Materiales CFM-Materials Physics
    Center MPC, centro mixto CSIC – UPV/EHU, 20018 Donostia-San
    Sebasti\'an, Spain}
  
  \address[4]{Department of Physics, University of Central Florida,
    Orlando, FL 32816, USA}
  
  \begin{abstract}
    For surface-mediated processes, such as on-surface synthesis,
    epitaxial growth and heterogeneous catalysis, a constant slope in
    the Arrhenius diagram of the corresponding rate of interest against 
    inverse temperature, $\log R$ {\it vs} $1/k_B T$, is traditionally
    interpreted as the existence of a bottleneck elementary reaction
    (or rate-determining step), whereby the constant slope (or
    apparent activation energy, $E_{app}^{R}$) reflects the value of
    the energy barrier for that reaction. Here, we show that a
    constant value of $E_{app}^{R}$ can be obtained even if control
    shifts from one elementary reaction to another. In fact, we show
    that $E_{app}^{R}$ is a weighted average and the leading
    elementary reaction will change with temperature while
    the actual energy contribution for every elementary reaction will
    contain, in addition to the traditional energy barrier, a
    configurational term directly related to the number of local
    configurations where that reaction can be performed.
    For this purpose, we consider kinetic Monte Carlo simulations of
    two-dimensional submonolayer growth at constant deposition flux,
    where the rate of interest is the tracer diffusivity. In particular, we
    focus on the study of the morphology, island density and
    diffusivity by including a large variety of single-atom,
    multi-atom and complete-island diffusion events for two specific
    metallic heteroepitaxial systems, namely, Cu on Ni(111) and Ni on
    Cu(111), as a function of coverage and temperature.
  \end{abstract}
\end{frontmatter}

\section{Introduction}
\label{sec:introduction}

Two dimensional (2D) materials have attracted interest due to their
superior properties and promising applications
\cite{doi:10.1021/acsnano.5b05040, lin_2d_2016, shen_frontiers_2017,
  ZHUIYKOV2017231, novoselov666, davila_germanene:_2014,
  grazianetti_two-dimensional_2016}. However, their future success
depends on the ability to achieve production in large amounts and with
high-quality, which directly relies on a better understanding of their
synthesis by a variety of surface-mediated processes
\cite{doi:10.1021/acsnano.5b05040}.  As an example of the many
techniques available, chemical vapour deposition (CVD) can be used to
grow a metal on top of the same metal (homoepitaxy) or on a different
metal (heteroepitaxy), which is also valid for the synthesis of novel
materials, such as graphene \cite{kim_large-scale_2009,
  huang_highly_2018, takesaki_highly_2016, seah_mechanisms_2014,
  tian_surface_2014, huet_role_2018,
  mueller_growing_2014,liu_synthesis_2011}. Traditionally, low energy
electron diffraction (LEED) \cite{davisson_scattering_1927,
  davisson_diffraction_1927, viefhaus_low-energy_1987}, field ion
microscopy (FIM) \cite{muller_field_1956,muller_resolution_1956},
scanning tunnelling microscopy (STM) \cite{binnig_scanning_1983} and
related microscopies have enabled the observation of single molecules
and atoms on the surface, thus providing specific insights regarding
the growth process \cite{antczak_surface_2010, knorr_long-range_2002,
  repp_substrate_2000, repp_site_2003, morgenstern_direct_2004}.

In general, surface-mediated growth uses a constant flux for each
vapour species, which is either adsorbed or thermally decomposed at
the surface. The resulting adparticles diffuse randomly, eventually
forming small clusters at random locations (nucleation), which
gradually evolve into larger islands through the attachment of other
diffusing adparticles (growth) until the islands eventually merge to
form a single 2D layer (coalescence). The quality of the 2D material
is affected markedly by the density and structure of the formed grain
boundaries, directly depending on the actual size and shape of the
islands (dendritic, compact, polygonal, ...), which ultimately depends
on the relative occurrence of the adsorption and diffusion events.  In
this context, the natural quantity describing the behaviour of the
system is the tracer diffusivity \cite{gosalvez2017}.

Due to the general character of the previous surface-mediated growth
mechanism, here we study submonolayer heteroepitaxial growth of
metals, in order to understand some of the global features, especially
the dominant contributions to the apparent activation energy of the
diffusivity.  In particular, we consider the growth of two
heteroepitaxial systems, namely, Cu on Ni(111) and Ni on Cu(111),
where the compact and stable (111) surfaces provide a small lattice
mismatch with respect to the growing 2D islands ($\sim 2.5 \%$), thus
facilitating surface diffusion and enabling the achievement of
concerted events, i.e. the diffusion of more than one adatom at once.
Although there are theoretical studies on (i) the diffusivity of a
single monomer of Cu (Ni) on Ni (Cu) \cite{acharya_diffusion_2017,
  onat_optimized_2014} and (ii) a more complete growth picture of
Cu/Ni(111) \cite{koschel_electronic_1999, koschel_growth_2000} and
Ni/Cu(111) \cite{boeglin_high_2002, pons_spontaneous_2002,
  mulazzi_structural_2006}, in this study we consider a large variety
of single-atom, multi-atom and complete-island diffusion events, with
the aim of obtaining a general picture on the relative importance of
concerted diffusion in two-dimensional material growth, applied to
metals.

The growth process is simulated by using the kinetic Monte Carlo (KMC)
method \cite{jansen_introduction_2012, gillespie_exact_1977}. As
opposed to molecular dynamics (MD), the KMC method avoids following
the motion of every possible atom contained in the system, simply
recognising that the important events---which modify the configuration
of the system and not just a few bond lengths---correspond to the
elementary reactions, each occurring after a certain wait time. In
this manner, every distinct elementary reaction (or rare event) is
assigned a different rate and time is discretised, with the resulting
time increment being much larger than in MD, thus enabling much longer
simulation times.  Based on a many body semi empirical embedded atom
model for the description of the interaction between the atoms
\cite{foiles_embedded-atom-method_1986}, the diffusion energy
barriers, $E_{\alpha}^{k}$, required to compute the rate of each
distinct diffusion event, are obtained by using the drag method
\cite{trushin_self-learning_2005, acharya_toward_2018}.  The
simulations allow switching on and off the concerted events as
desired, thus enabling the study of their relative importance against
other events, including their contribution to the apparent activation
energy, as well as their effect on the morphology of the generated
islands.

In this respect, the actual shapes of the islands typically look
different from one simulation to another, due to the stochastic nature
of the KMC method. However, the simulations performed using the same
rates (= rate constants) display common features and simple visual
inspection will conclude that the islands are equivalent in some
manner. In the present study, this is demonstrated quantitatively by
performing a power spectral density (PSD) analysis
\cite{persson2005nature, elson1995calculation, czifra2009sensitivity,
  Ferrando2014}. Here, an image of the surface is associated with a 2D
map, where each location represents a harmonic frequency and the
displayed value represents the squared sum of the real and imaginary
amplitudes for that harmonic component (i.e., the power for that
frequency). Low frequencies are correlated to large structures, such
as the overall shape of the islands, while large frequencies are
related to small features, such as the structure of the perimeter. In
this manner, the PSD analysis enables comparing different/similar
surfaces with stochastic variations. In fact, two PSD maps can be
considered equivalent when their point-to-point difference produces
noise (= stochastic fluctuations) around the 0 value all over the
resulting difference map. On the contrary, when two PSDs differ
structurally, their difference map displays distinctive patterns,
clearly deviating from random fluctuations around the 0 value. This
enables determining the effect on the morphology of the generated
islands due to switching on or off certain diffusion events.

In order to describe the dominant contributions to the apparent
activation energy of the tracer diffusivity, section
\ref{sec:diffusivity} presents the direct relation that exists between
the diffusivity and the total diffusion rate ($=$ total hop rate). The
total diffusion rate, in turn, depends directly on the multiplicities
of the different diffusion events (i.e., the multiplicity is the
actual number of locations where each distinct diffusion event can be
performed in a given snapshot of the surface).  This is followed by a
description of all the different diffusion events considered in the
study, including single-atom, multi-atom and complete-island diffusion
events as well as their energy barriers in sections
\ref{sec:identification}-\ref{sec:EnergyBarriers}. For clarity,
section \ref{sec:theory_R} provides a detailed description of the
total diffusion rate, total adsorption rate and total rate as well as
their time and ensemble averages in terms of the corresponding
multiplicities, and section \ref{sec:ae} shows, as a result, that the
apparent activation energy of any of the total rates depends on the
multiplicities and, thus, the apparent activation energy of the
diffusivity as well. Finally, section \ref{sec:kmc} culminates the
presentation of the theoretical and computational aspects of the study
by describing the most salient features of the implemented KMC
method. In addition, sections
\ref{sec:IslandDensity}-\ref{sec:Results_ae} present the results of
the study, comparing the temperature dependence of the island density,
morphology, total rates and their apparent activation energy for the
two chosen systems, namely, Cu on Ni(111) and Ni on Cu(111).  Finally,
section \ref{sec:conclusions} summarises the conclusions of the study.


\section{Computational details and theoretical aspects}
\subsection{Tracer diffusivity in surface-mediated growth}
\label{sec:diffusivity}
The natural quantity describing submonolayer growth under a constant flux of adparticles is the tracer diffusivity
\cite{gosalvez2017}:
\begin{equation}
  D_T = \frac{1}{2 \delta \langle \hat{n}_{a} \rangle } \frac{\langle \mathcal{\hat{R}}^2 \rangle}{t} , \label{eq:D_T_0a}
\end{equation}
where the hat symbol ($\; \hat{ } \;$) denotes the value of a
time-dependent variable at time $t$, $\delta$ is the dimensionality
($=2$ for diffusion on a surface), $\hat{n}_{a}$ is the number of
adsorbed particles,
$\mathcal{\hat{R}}^2 = \Sigma_{i=1}^{\hat{n}_{a}} |
\mathbf{\hat{x}}_{i}-\mathbf{x}_{i}^* |^2$ is the total squared
distance travelled by the adparticles, with $\mathbf{x}_{i}^*$
denoting the position of adparticle $i$ when it was adsorbed, and
$\langle \hat{X} \rangle$ is the ensemble average of $\hat{X}$ over
$K$ samples in the limit of large $K$.  In paticular, the diffusivity
can be re-written as \cite{gosalvez2017}:
\begin{align}
  D_T & = \frac{l^2}{2 \delta \theta } f_T R_d \; ,  \label{eq:D_T_0b} \\  
  R_d & = \sum_{ \alpha \in \{ d \} } M_\alpha k_\alpha \label{eq:D_T_0c} ,
\end{align}
where $l$ is the hop distance between adjacent sites,
$\theta = \langle \hat{\theta} \rangle$ is the ensemble average of the
coverage, $\hat{\theta} = \hat{n}_{a} / L_{x}L_{y}$, with $L_{x}L_{y}$
being the total number of adsorption sites,
$R_d=\langle \overline{ \hat{R_d} } \rangle$ is the time and
ensemble average of the total diffusion rate per site, $\hat{R_d}$,
where
$\overline{\hat{X}} = \frac{\int \hat{X} dt}{\int dt} =
\frac{\Sigma_k \hat{X}_k \Delta t_k}{\Sigma_k \Delta t_k}$
is the time average of $\hat{X}$, and $f_T$ is the correlation
factor, which accounts for memory effects between consecutive hops at
finite coverages, {\it e.g.} hopping from site $i$ to site $j$ leaves
site $i$ empty and, thus, at finite coverage the adparticle has a
higher chance of returning to $i$ \cite{gosalvez2017}.  Finally,
equation \ref{eq:D_T_0c} is the time and ensemble average of:
$\hat{R_d} = \Sigma_{ \alpha \in \{ d \} } \hat{M}_\alpha
k_\alpha$, which defines the total diffusion rate per site.
The summation is over the collection of all distinct
diffusion events $\{ d \}$, $k_\alpha$ is the rate of diffusion
event $\alpha$ (referred to as the {\it rate constant} or {\it
  specific rate} in chemical kinetics) and
$M_\alpha=\langle \overline{ \hat{M_\alpha} } \rangle$ is the time
and ensemble average of the multiplicity,
$\hat{M}_\alpha = \hat{m}_\alpha / L_{x}L_{y}$, with
$\hat{m}_\alpha$ being the number of locations where diffusion event
$\alpha$ can be performed in a given snapshot of the surface.  Here,
$k_\alpha$, is determined by using transition state theory (TST):
\begin{equation}
  k_\alpha = k_{0} e^{-E_\alpha^k / k_B T} ,
  \label{eq:tst}
\end{equation}
where $k_B$ is the Boltzmann constant, $T$ is the temperature,
$E_\alpha^k$ is the energy barrier for the diffusion event and
$k_{0}$ is the attempt frequency, which depends weakly on temperature
and is usually assigned the value of $10^{13}$ Hz.

Equation \ref{eq:D_T_0b} means that the apparent activation energy of
the diffusivity is essentially given by that of the total diffusion
rate. In fact, using equation \ref{eq:D_T_0c} in equation
\ref{eq:D_T_0b} and denoting the inverse temperature as
$\beta = 1 / k_{B} T$ while considering that the coverage is
independent of temperature (due to the constant flux), the apparent
activation energy of the diffusivity,
$E_{app}^{D_T} = -\frac{ \partial \log D_T }{ \partial \beta }$,
is easily determined \cite{gosalvez2017}:
\begin{align}
E_{app}^{D_T} & = E^f + E_{app}^{R_d}, \label{EappD_T_0} \\
E_{app}^{R_d} & = \sum_{\alpha \in \{ d \}} \omega_\alpha^{R_d} (E_\alpha^k + E_\alpha^{M}) , \label{EappR_d_0} \\
\omega_\alpha^{R_d} & = \frac{ M_\alpha k_\alpha }{ R_d }= \frac{ M_\alpha k_\alpha }{ \Sigma_{ \alpha' \in \{ d \} } M_{\alpha'} k_{\alpha'} } \; , \; \alpha \in \{ d \} . \label{omega_definition_M}
\end{align}
Here, $E^f = -\frac{ \partial \log f_T }{ \partial \beta }$,
$E_\alpha^k = -\frac{ \partial \log k_\alpha }{ \partial \beta }$ and
$E_\alpha^{M} = -\frac{ \partial \log M_\alpha }{ \partial \beta }$
are the contributions from the correlation factor ($f_T$), the rate of
diffusion event $\alpha$ ($k_\alpha$) and the corresponding
multiplicity ($M_\alpha$), respectively, and the weight
$\omega_\alpha^{R_d}$ is the \emph{probability} of observing diffusion
event $\alpha$ amongst all distinct diffusion events. Indeed, the
\emph{event probabilities} of equation \ref{omega_definition_M} are
very useful, providing a complete picture of the undergoing
competition between the different diffusion events, directly
indicating which events dominate and which are essentially
irrelevant. Typically, the contribution from the correlation factor is
small ($E^f \approx 0$) \cite{gosalvez2017}. Thus, equation
\ref{EappD_T_0} shows that the temperature dependence of the
diffusivity is essentially given by that of the total diffusion rate:
$E_{app}^{D_T} = E^f + E_{app}^{R_d} \approx E_{app}^{R_d} =
\Sigma_{\alpha \in \{ d \}} \omega_\alpha^{R_d} (E_\alpha^k +
E_\alpha^{M})$. In this manner, we focus below on the analysis 
of the total diffusion rate.

\subsection{Identification of diffusion events}
\label{sec:identification}
The diffusion of adsorbates on a substrate is an essential part of
film growth. In general, a diffusion event may consist in a
single-atom hop (single-atom diffusion), a complete-island hop
(concerted island diffusion) or a multi-atom hop at the perimeter of a
compact island (concerted multi-atom diffusion)
\cite{wang_surface_1998, schlos_ser_kinetics_2000,
  kellogg_oscillatory_1994, zhang_atomistic_1997, wang_diffusion_1997,
  wu_island_2010, signor_misfit-dislocation-mediated_2010}.  Here, an
island is considered as a structure where each atom is connected with
at least one nearest neighbour.  Based on an extensive study of
post-adsorption diffusion kinetics of small islands of Cu/Ni(111) and
Ni/Cu(111) as a function of the island size
\cite{acharya_diffusion_2017, acharya_toward_2018}, it is concluded
that, in addition to single-atom diffusion, the most executed
diffusive events correspond to concerted diffusion of complete islands
with size up to eight atoms and concerted two-atom diffusion along the
step-edge of compact islands.  Thus, the current study focuses on
including these particular diffusion events.

\begin{figure}[htb!]
  \centering
  \def\svgwidth{\columnwidth}
\executeiffilenewer{sourceTypes.svg}{sourceTypes.pdf}%
{inkscape             -z             -D             --file=sourceTypes.svg             %
--export-pdf=sourceTypes.pdf             --export-latex}%
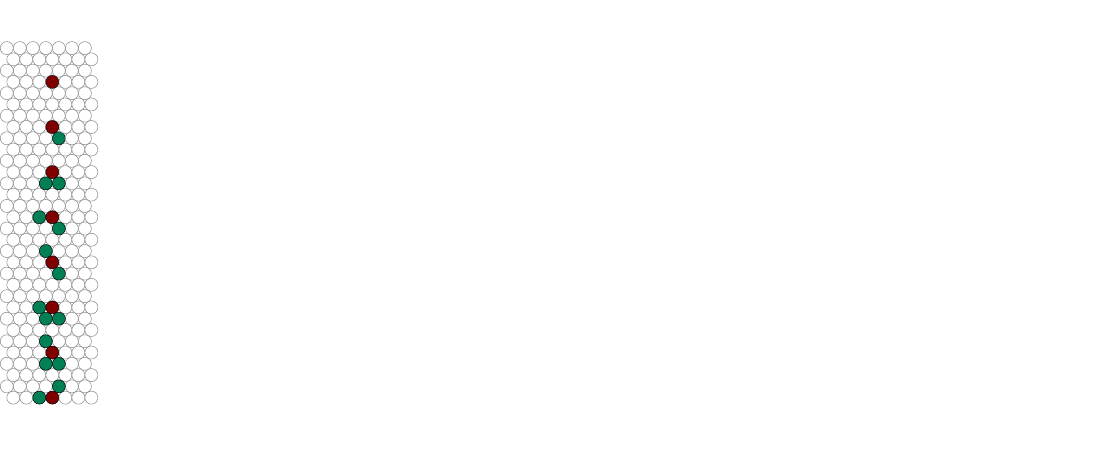%

  \caption{Site types considered in this study, indicated for the
    central site (in red). The type is a unique combination of class
    and subclass. The class directly indicates the number of occupied
    nearest neighbour sites (in green). The subclass is a label used
    to distinguish between different geometrical arrangements of the
    occupied neighbours.}
  \label{fig:types}
\end{figure}

The crystallographic structure of the fcc(111) surface under
consideration in this study is described by a triangular lattice,
where every node represents an adsorption site. Any site is assigned a
type (the unique combination of a class and a subclass), regardless of
being occupied by an atom or not. The site class directly indicates
the number of occupied nearest neighbour sites, and the subclass is
simply a label that allows distinguishing between the different
geometrical arrangements of the occupied neighbour sites.  As shown in
figure \ref{fig:types}, we consider 7 site classes (from 0 to 6) and a
maximum of 3 subclasses (from 0 to 2), leading to a total of 13 site
types. The subclass is always 0 for classes 0, 1, 5 and 6, while it is
arbitrarily and consistently assigned the value 0, 1 or 2 for classes
2, 3 and 4, depending on the geometrical arrangement.

\begin{figure}[htb!]
  \centering
  \def\svgwidth{\columnwidth}
  {\scriptsize
\executeiffilenewer{types.svg}{types.pdf}%
{inkscape             -z             -D             --file=types.svg             %
--export-pdf=types.pdf             --export-latex}%
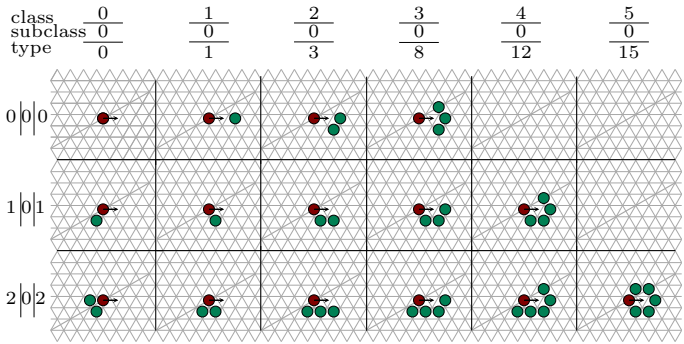%

    }
  \caption{An extract of the total transition table for single-atom
    diffusion on a triangular lattice, showing a few representative
    origin site types (rows) and destination site types (columns). In
    all cases the red particle moves one lattice site to the
    right. The complete table appears in figure \ref{fig:all_energies}
    of the appendix.}
  \label{fig:some_energies}
\end{figure}

For any particular diffusion event, the destination site is assigned the
site type by considering that the atom has already hopped on it. Thus,
for a destination site, the class and subclass are determined by
considering the origin site as being empty. This is shown in figure
\ref{fig:some_energies}. For the first column, the destination site
will have no occupied neighbours and, thus, the type is 0 (class 0,
subclass 0); for the second column, the destination site will have one
occupied neighbour and, therefore, the type is 1 (class 1, subclass
0); and so on. Note that, in practice, there are only 12 origin types
(0 to 11), since diffusion is impossible from type 12 (class 6,
subtype 0). Furthermore, regarding the destination types, we consider
some additional cases in order to take into account detachment
events. Here, detachment means that the destination site has no
neighbours in common with the origin site. This leads to a total of 16
destination types, as shown in figure \ref{fig:all_energies} of the
appendix. Consequently, we work with a transition table of 12 $\times$
16 entries, where the rows correspond to the origin site types and the
columns to the destination site types.

In order to assign a type to a diffusion event, we use the origin and
destination types of the involved sites. Some representative
single-atom diffusion events are shown in figure
\ref{fig:some_energies}, where the atom at the origin site (in red) is
moved one site to the right in each of the 15 examples.  In this
manner, monomer diffusion is described by a hop from type 0 to type 0
($D[0,0]\rightarrow [0,0]$), while edge diffusion is described by a
hop from type 2 to type 3 ($D[2,0]\rightarrow [2,0]$).  Note that
some transitions are physically impossible and, thus, not displayed.
Similarly, we do not display the symmetry equivalent transitions for
the other five hop directions (there are 6 possible directions in a
triangular lattice), since all directions are treated identically.

In addition to single atom hops, we include concerted island diffusion
up to 8 atoms, where all the atoms belonging to the island move
together in one of the six directions, independently of the island
shape.  For the calculation of the corresponding energy barrier (see
section \ref{sec:EnergyBarriers}) the most compact shapes are used, as
shown in figure \ref{fig:islandGeom}. For instance, this means that
all different trimer shapes have the same rate to move in any
direction. Finally, we also include concerted two-atom diffusion along
the perimeter of compact islands according to the four different event
types shown in figure \ref{fig:multiAtomGeom}. Overall, we consider
118 different diffusion event types: 107 single-atom diffusions
(figure \ref{fig:all_energies} of the appendix), 7 complete-island
moves (figure \ref{fig:islandGeom}) and 4 multi-atom hops (figure
\ref{fig:multiAtomGeom}).
\begin{figure}[htb!]
  \centering
  \def\svgwidth{0.7\columnwidth}
  {\scriptsize
\executeiffilenewer{islandDiffusionGeometries.svg}{islandDiffusionGeometries.pdf}%
{inkscape             -z             -D             --file=islandDiffusionGeometries.svg             %
--export-pdf=islandDiffusionGeometries.pdf             --export-latex}%
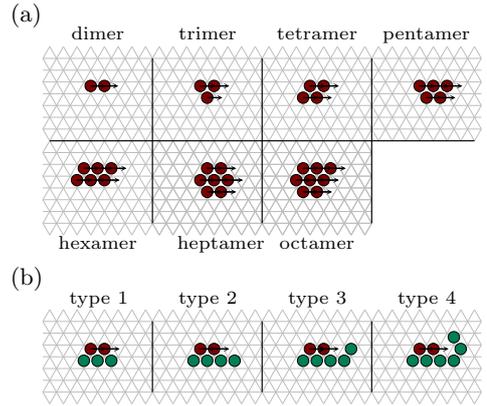%

  }
  \caption{ (\subref{fig:islandGeom}) Compact islands containing up to
    8 atoms, used for the calculation of the diffusion energy barrier
    for concerted island diffusion.  (\subref{fig:multiAtomGeom})
    Concerted diffusion events for two-atoms (in red) at the perimeter
    of an island (in green), classified according to the destination
    site type for the rightmost diffusing atom.}
  \label{fig:islandsAndMultiAtom}
\end{figure}

\subsection{Calculation of energy barriers and rates} 
\label{sec:EnergyBarriers}

For each identified diffusion event, the activation energy barrier
($E_{\alpha}^{k}$) is calculated by using the drag method while
describing the interaction between the atoms with a many-body
semi-empirical embedded atom model (EAM)
\cite{foiles_embedded-atom-method_1986}. The EAM + drag combination
produces qualitative and semi-quantitative results with minor errors
for metallic systems when compared with ab initio energetics,
including island diffusion barriers on fcc(111)
\cite{acharya_diffusion_2017, trushin_self-learning_2005}. For this
study, the substrate consists on five fcc(111) layers with
$16\times16 = 256$ atoms per layer, where the two bottom layers are
kept frozen (to mimic the bulk) while the three top layers are
allowed to relax. For each diffusion event, the required adatoms are
placed on the surface on the desired initial configuration and the
system is relaxed (standard MD cooling with velocity updates using the
leap-frog algorithm) until the energy change between successive
updates is less than $10^{-4}$ eV, taking the corresponding minimised
energy as reference for the calculation of $E_{\alpha}^{k}$.

In order to determine the energy barrier for a single-atom diffusion
event, the chosen adatom is gradually dragged in steps of 0.05
\AA~along the reaction coordinate, whose direction is re-defined at
every step as the vector from the relaxed location of the adatom to
the aimed location in the final configuration. At every step,
relaxation is allowed for the dragged adatom along the plane
perpendicular to the current direction of the reaction coordinate
while keeping fixed all other adatoms, the two bottom layers and the
reaction coordinate, until the energy difference is less than
$10^{-3}$ eV (1 meV) or the relaxed adatom is 0.05~\AA~from the aimed
location. The maximum energy point in the energy profile of the
minimum energy path represents the saddle point and its energy
difference from the reference energy gives $E_{\alpha}^{k}$ for the
diffusion event. For multi-atom and concerted diffusion events, the
same procedure is applied to the adatoms under consideration. See
Ref. \cite{acharya_toward_2018} for further details.

The computed energy barriers are displayed in tables \ref{tb:cuni} and
\ref{tb:nicu} of the appendix. The barriers for monomer diffusion are
within the expected range, compared with the literature
\cite{antczak_surface_2010}. Our diffusion barrier of 52 meV for Cu on
Ni(111) is virtually the same as the previously reported value of 50
meV \cite{kim2007transition}. For diffusion of Ni on Cu(111), we
obtain 31 meV, which is about 2/3 of the value (45 meV) reported in
Ref. \cite{kim2007transition}.  Once the energy barrier $E_{\alpha}^{k}$
has been obtained for diffusion event $\alpha$, the corresponding
diffusion rate ($k_{\alpha}$) is computed by using equation
\ref{eq:tst}.

\subsection{Total rates and the probability to observe an event}
\label{sec:theory_R}
In order to model the adsorption and diffusion of adatoms of Ni on
Cu(111) and Cu on Ni(111), we consider a two-dimensional lattice of
adsorption sites under a typical constant deposition flux
\cite{gosalvez2017, Venables1984, Amar1994, Ratsch2003,
  Korner2012}. For each of the two systems, the substrate is treated
as a two-dimensional triangular lattice, where atoms from the
surrounding environment are deposited randomly (on the empty sites)
while previously adsorbed adatoms are able to diffusive according to
the particular diffusion events considered in section
\ref{sec:EnergyBarriers} for single atoms, multiple atoms and complete
islands. Desorption events are neglected, due to their extremely low
rate in these systems. A constant deposition flux, $F$, is considered
at any temperature and, thus, a temperature-independent adsorption
rate per site is used, $k_{a}=F$, independently of the occupation
state of the neighborhood of the empty site where adsorption may
occur. For diffusion, the temperature-dependent diffusion rate for
event type $\alpha$, $k_\alpha$, is given by equation \ref{eq:tst}.
Below, instantaneous values of time-dependent variables are indicated
by using the hat symbol ($\; \hat{ } \;$).  The theoretical
presentation provided below follows closely that given in
Ref. \cite{gosalvez_arxiv}.

At any given time $t$, the total diffusion rate is:
\begin{equation}
  \hat{r}_d = \sum_{\alpha \in \{ d \} } \hat{m}_\alpha k_\alpha, \label{Rh_definition}
\end{equation}
where $\hat{m}_\alpha$ is the multiplicity for diffusion event
$\alpha$, i.e.\ the number of locations where that particular
diffusion event can be performed on the current configuration of the
surface, $k_\alpha$ is the corresponding diffusion rate, and
$\{ d \}$ is the complete collection of distinct diffusion
events.
Similarly, the total adsorption rate is:
\begin{align}
  \hat{r}_{a} & = \sum_{\alpha \in \{ a \} } \hat{m}_\alpha k_\alpha, \label{Ra_definition00} \\
              & = \hat{m}_{a} k_{a} , \label{Ra_definition01} \\
              & = (1- \hat{\theta} )L_{x}L_{y} k_{a} = L_{x}L_{y} e^{-k_{a}t} k_{a} , \label{Ra_definition}
\end{align}
where $\{ a \}$ is the collection of distinct adsorption events, and
$\hat{m}_\alpha$ ($k_\alpha$) is the corresponding multiplicity
(adsorption rate).  Since we consider only one adsorption event type,
the summation in equation \ref{Ra_definition00} is reduced to a single
term, as indicated in equation \ref{Ra_definition01}. Here, $k_{a}=F$
and $\hat{m}_{a}$ is the corresponding multiplicity, i.e.\ the total
number of empty sites. Note that
$\hat{m}_{a}=(1 - \hat{\theta} ) L_{x}L_{y}$ in equation
\ref{Ra_definition}, where $\hat{\theta} = \hat{n}_{a} / L_{x}L_{y} $
designates the coverage, with $\hat{n}_{a}$ being the total number of
adsorbed atoms up to the current time (i.e.\ the total number of
adsorptions events) and $L_{x}L_{y}$ the total number of adsorption
sites (before adsorption of any atom). Due to the constant deposition
flux, the coverage increases with time according to the equation:
$\frac{d \hat{\theta} }{dt} = k_{a}(1- \hat{\theta} )$, which is
directly integrated to give: $\hat{\theta} = 1 - e^{-k_{a}t}$. Thus,
$\hat{m}_{a}=L_{x}L_{y}e^{-k_{a}t}$ independently of the temperature.

Finally, since both diffusion and adsorption events may
occur, we consider the total rate:
\begin{align}
  \hat{r} & = \sum_{\alpha \in \{ e \}} \hat{m}_\alpha k_\alpha, \label{Re_definition01} \\
   & = \sum_{\alpha \in \{ d \}} \hat{m}_\alpha k_\alpha +  \sum_{\alpha \in \{ a \}} \hat{m}_\alpha k_\alpha , \label{Re_definition02} \\
   & = \sum_{\alpha \in \{ d \}} \hat{m}_\alpha k_\alpha +  \hat{m}_{a} k_{a} , \label{Re_definition03} \\
   & = \hat{r}_d + \hat{r}_{a}, \label{Re_definition}
\end{align}
where $\{ e \}$ is the collection of all distinct event types
(diffusion and adsorption). The total number of performed events is:
$\hat{n} = \hat{n}_d + \hat{n}_{a}$, where $\hat{n}_d$ is the
total number of performed hops and $\hat{n}_{a}$ is the total number
of executed adsorptions (as defined above).

Dividing by the total number of sites, $L_{x}L_{y}$, we also define
the \emph{total diffusion rate per site},
$\hat{R}_{d} = \hat{r}_{d} / L_{x}L_{y}$, the \emph{total adsorption
  rate per site}, $\hat{R}_{a} = \hat{r}_{a} / L_{x}L_{y}$, the
\emph{total rate per site}, $\hat{R} = \hat{r} / L_{x}L_{y}$, and the
\emph{multiplicity per site},
$\hat{M}_\alpha = \hat{m}_\alpha / L_{x}L_{y}$. For simplicity, both
$\hat{m}_\alpha$ and $\hat{M}_\alpha$ are referred to as the
\emph{multiplicity}, although $\hat{M}_\alpha$ should be understood as
a multiplicity density or relative abundance or
concentration. Similarly, $\hat{R}_{d}$, $\hat{R}_{a}$ and $\hat{R}$
may be referred to as the \emph{total rates}, thus obviating their
per-site character.  The total diffusion rates, $\hat{r}_{d}$ and
$\hat{R}_{d}$, are important, since the tracer diffusivity, $D_{T}$,
is proportional to their average, as shown in equation
\ref{eq:D_T_0b}. Similarly, $\hat{r}$ and $\hat{R}$ are also
important, since the inverse of $\hat{r}$ provides a natural measure
of the time increment: $\Delta t = -\log(u) / \hat{r_{}} $, where
$u \in (0,1]$ is a uniform random number.  By definition, $\hat{r}$ is
equal to the number of performed events per unit time,
$\hat{r} = \frac{ d\hat{n} }{ dt }$, and thus,
$\hat{r} = \frac{ 1 }{ \Delta t }$, since exactly one event occurs in
every time step. With a mean value of $1$, the positive factor
$-\log(u)$ enforces the correct Poisson distribution for the time
steps.

Making the observation that $\hat{r}_{d}$ is equal to the number of
performed diffusion events per unit time,
$\hat{r}_{d} = \frac{ d\hat{n}_{d} }{ dt }$ (similar to
$\hat{r} = \frac{ d\hat{n} }{ dt }$), the time average of
$\hat{r}_{d}$ for any desired period is written exactly as the total
number of performed diffusion events, $\hat{n}_{d}$, divided by the
elapsed time, $t$ (and similarly for $\hat{r}_{a}$ and $\hat{r}$):

\begin{eqnarray}
  \overline{ \hat{r}_{d} } = \frac{ \int \hat{r}_{d} dt }{ \int dt } = \frac{ \int \frac{ d\hat{n}_{d} }{ dt } dt }{ t } = \frac{ \int d\hat{n}_{d} }{ t } = \frac{\hat{n}_{d}}{t} , & \label{eq_rd_t_av} \\
  \overline{ \hat{r}_a } =  \frac{\hat{n}_{a}}{t} , & \label{eq_ra_t_av} \\
  \overline{ \hat{r} } =  \frac{\hat{n}}{t} . & \label{eq_r_t_av} 
\end{eqnarray}

Carrying out the ensemble average in
Eqs. \ref{eq_rd_t_av}-\ref{eq_r_t_av} and dividing by $L_{x}L_{y}$
gives: \vspace{-2mm}
\begin{align}
R_d = & \frac{N_d }{ \tau }, \label{eq:AverageRh1} \\
R_{a} = & \frac{N_{a} }{ \tau } , \label{eq:AverageRa1} \\
R = & \frac{N }{ \tau }, \label{eq:AverageRe1}
\end{align}

where $R_d = \langle \overline{ \hat{R}_d } \rangle$,
$R_{a} = \langle \overline{ \hat{R}_{a} } \rangle$ and
$R = \langle \overline{ \hat{R} } \rangle$ are the \emph{average total
  rates per site} (for diffusion, adsorption and all events,
respectively), while $N_d = \langle \hat{n}_d \rangle / L_{x}L_{y}$,
$N_{a} = \langle \hat{n}_{a} \rangle / L_{x}L_{y}$, and
$N = \langle \hat{n} \rangle / L_{x}L_{y}$ specify the ensemble
averages of the numbers of performed events per site (for diffusion,
adsorption and all events, respectively), and
$\tau = \langle t \rangle $ is the ensemble average of the elapsed
time.  On the other hand, performing the time and ensemble averages on
equations \ref{Ra_definition01}, \ref{Rh_definition} and
\ref{Re_definition01} and dividing by the total number of adsorption
sites gives:
\begin{align}
  R_d & = \sum_{\alpha \in \{ d \} } M_\alpha k_\alpha , \label{eq:AverageRh2} \\
  R_{a} & = M_{a}  k_{a} = (1- \theta) F, \label{eq:AverageRa2} \\
  R & = R_d + R_{a} = \sum_{\alpha \in \{ e \} } M_\alpha  k_\alpha , \label{eq:AverageRe2}
\end{align}
where
$M_\alpha = \langle \overline{ \hat{m}_\alpha } \rangle / L_{x}L_{y} $
and
$M_{a} = \langle \overline{ \hat{m}_{a} } \rangle / L_{x}L_{y} = 1-
\theta$ are the corresponding time and ensemble averages of the
multiplicities per site. Here, $\theta = \langle \hat{\theta} \rangle$
is the ensemble average of the coverage. Equations
\ref{eq:AverageRh1}-\ref{eq:AverageRe1} and
\ref{eq:AverageRh2}-\ref{eq:AverageRe2} are very important for this
study, since they provide two alternative expressions to determine the
same quantities ($R_d$, $R_{a}$ and $R$). In addition to ensuring the
correct determination of each quantity, the equations provide a way to
describe the particular contributions that make up any specific value
of their apparent activation energy.

Finally, we note that, in addition to the relation to the time
increment, the average total rate per site, $R$, is very important,
since it is used in the definition of the \emph{probability} of
observing event $\alpha$ amongst all distinct events (diffusion and
adsorption):
\begin{equation}
  \omega_\alpha^{R} = \frac{ M_\alpha k_\alpha }{ R } = \frac{ M_\alpha k_\alpha }{ \Sigma_{\alpha' \in \{ e \} } M_{\alpha'} k_{\alpha'} } \; , \; \alpha \in \{ e \}.
  \label{omega_definition}
\end{equation}
Note the difference with respect to $\omega_{\alpha}^{R_{d}}$ in
equation \ref{omega_definition_M}. Although both quantities are
probabilities, their meaning is with respect to the collection of
events considered in the denominator, thus justifying the superindex
$R$ or $R_{d}$, respectively. The \emph{event probabilities} of
equation \ref{omega_definition} directly indicate which events
dominate the overall process, considering both adsorption and
diffusion events. The \emph{event probabilities} of equation
\ref{omega_definition_M} indicate which diffusion events dominate with
respect to all distinct diffusion events. Due to the link of $R$ to
the overall event probabilities and of $R_{d}$ to the tracer
diffusivity, we focus on the analysis of the temperature dependence of
both quantities.

\subsection{Apparent activation energy}
\label{sec:ae}

For an Arrhenius plot of the average total rate per site [where
$\log( R )$ is drawn against inverse temperature,
$\beta = 1 / k_B T$], the apparent activation energy is defined as:
\begin{align}
  E_{app}^{R}  & = - \frac{ \partial   \log R }{ \partial   \beta } , \label{EappR_0} \\
               & = - \frac{ 1 } { R } \frac{ \partial R }{ \partial   \beta } , \label{EappR_1} \\
               & = - \frac{ 1 } { \sum_{\alpha \in \{ e \}} M_\alpha k_\alpha } \frac{ \partial (  \sum_{\alpha \in \{ e \}} M_\alpha k_\alpha ) }{ \partial   \beta } , \label{EappR_2}
\end{align}
where equation \ref{eq:AverageRe2} has been used to write equation
\ref{EappR_2}. Since the multiplicities, $\hat{m}_\alpha$, depend on
the actual values of the event rates, $k_\alpha$, the average
multiplicities per site, $M_\alpha$, are functions of
temperature. Using
$k_\alpha = k_\alpha^{0} e^{-E_\alpha^{k} \beta}$ and
$E_\alpha^{M} = - \frac{ \partial \log M_\alpha }{ \partial \beta
}$, and applying the chain rule to
$\sum_{\alpha \in \{ e \}} M_\alpha k_\alpha$ easily leads to:
\begin{equation}
  E_{app}^{R} = \sum_{\alpha \in \{ e \}}  \epsilon_\alpha^{R}  ,  \; \: \rm{with} \; \; \epsilon_\alpha^{R} = \omega_\alpha^{R} (E_\alpha^{k} + E_\alpha^{M}) , \label{EappR_3}
\end{equation}
where the weight $\omega_\alpha^{R}$ is the probability to observe
event $\alpha$, as given in equation \ref{omega_definition}. Note
that, in general, the additional term
$E_\alpha^{k^{0}} = - \frac{ \partial \log(k_\alpha^{0}) }{
  \partial \beta }$ should be added to $E_\alpha^{k}+E_\alpha^{M}$
in equation \ref{EappR_3}. However, $E_\alpha^{k^{0}}$ is zero in
this study, since the prefactors $k_\alpha^{0}$ are
temperature-independent.

According to equation \ref{EappR_3}, the contribution of event type
$\alpha$ to the apparent activation energy is
$\epsilon_\alpha^{R} = \omega_\alpha^{R} (E_\alpha^{k} +
E_\alpha^{M})$. Simply speaking, $\omega_\alpha^R$ provides the
relative importance of event type $\alpha$ as compared to all other
events. Inside the bracket, the microscopic activation energy for the
event type, $E_\alpha^{k}$, is modified by a configurational
contribution, $E_\alpha^{M}$, which describes how the multiplicity of
that event type changes with temperature.  Note that $E_{\alpha}^{M}$
is unbounded and can be positive, negative or zero, depending on the
actual increase, decrease or constancy of $M_{\alpha}$ locally with
respect to temperature.  Thus, when the overall process is dominated
by a single event type or \emph{rate determining step}, say $\lambda$,
then $\omega_\lambda^R \approx 1$ and $\omega_\alpha^R \approx 0$ for
all other event types, and we have:
$E_{app}^{R} = E_\lambda^{k} + E_\lambda^{M}$. In this manner, the
apparent activation energy, $E_{app}^{R}$, differs from the activation
barrier of the rate determining step, $E_\lambda^{k}$, due to the
change in the number of locations where that particular event can be
performed on the surface as a function of temperature,
$E_\lambda^{M}$.

The apparent activation energy $E_{app}^{R_{d}}$ of the total diffusion rate, $R_{d}$, is
obtained similarly and the result is given in equation
\ref{EappR_d_0}. Thus, the apparent activation energies for $R$ and
$R_{d}$ have the same functional dependence, only differing in the
actual collection of considered events (both diffusion and adsorption
events for $R$, and only the diffusion events for $R_{d}$) and,
correspondingly, the value of the weight, i.e.\ the probability with
respect to the other considered events.

\subsection{Kinetic Monte Carlo}
\label{sec:kmc}

\begin{figure}[!ht]
  \centering
  \def\svgwidth{0.6\columnwidth}
\executeiffilenewer{flowchart.svg}{flowchart.pdf}%
{inkscape             -z             -D             --file=flowchart.svg             %
--export-pdf=flowchart.pdf             --export-latex}%
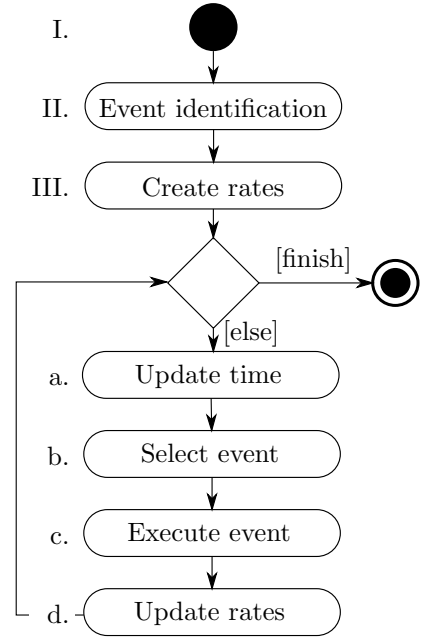%

  \caption{Unified modelling language (UML) activity diagram of a generic KMC algorithm.}
  \label{fig:flowchart}
\end{figure}

For the actual simulations, we use the standard, rejection-free,
time-dependent implementation of the kinetic Monte Carlo (KMC) method
with periodic boundary conditions \cite{gosalvez2017,
  jansen_introduction_2012, trushin_self-learning_2005,
  acharya_diffusion_2017, Ferrando2014}. A flowchart of the KMC
procedure is presented in figure \ref{fig:flowchart}. Regarding the
central rhombus in figure \ref{fig:flowchart}, a threefold termination
criterion is used, based on surpassing any of the maximum values
specified by the user for (i) the coverage $\hat{\theta}$, (ii) the
simulated time $t$, and (iii) the total number of simulated events
$\hat{n} = \hat{n}_d + \hat{n}_{a}$. In order to initiate the
simulation (and keep it going) the fundamental ingredients are the
combination of a specific geometry (figure \ref{fig:flowchart}I, here
a triangular lattice), a complete list of possible events (figure
\ref{fig:flowchart}II.) and their rates (figure
\ref{fig:flowchart}III.). Although nothing prevents starting from an
arbitrary coverage, in this study the initial configuration is always
an empty surface (no adatoms).

Initially, the stop criteria are not met and the main loop starts by
updating the simulated time (figure \ref{fig:flowchart}a.). This is
done by adding the time increment $ \Delta t = -\log(u)/\hat{r}$ to
the current value of $t$, as indicated in section
\ref{sec:theory_R}. Continuing with the algorithm, a random number is
used to select the next event that will be executed (figure
\ref{fig:flowchart}b.). This is done by randomly choosing one rate
among all the current rates, i.e.\ among all the diffusion and
adsorption events that are currently possible. The next step in the
algorithm is to execute the selected event (figure
\ref{fig:flowchart}c.). From a computational perspective, adsorption
implies adding an atom to an empty site while, in general, diffusion
requires removing several atoms from the occupied initial sites and
adding them to the empty final sites. As a result, the adsorption and
diffusion rates need to be updated for the involved sites as well as
their neighbours (figure \ref{fig:flowchart}d.), adding and deleting
available events too. The main loop finishes here and it is repeated
until a stop criterion is met.

For this study we have used the software known as ``Morphokinetics'',
written in object-oriented Java language and developed at the Donostia
International Physics Center. Based on the KMC method, Morphokinetcs
enables simulating various surface-mediated processes, including
anisotropic etching (removal of material from the substrate),
heterogeneous catalysis (reactions on the substrate) and 2D monolayer
growth (deposition of material on the substrate). The source code is
freely available at the GitHub repository
\footnote{\url{https://github.com/dipc-cc/Morphokinetics.git}}, with a
free license GNU General Public License version 3 or any later
version, which means that users have freedom to run, study,
redistribute and improve the program.

In the simulations, the adsorption flux is fixed to $1.5 \times 10^4$
ML/s and the temperature is varied from 23 to 1000 K for both systems
under study. We use just one value of the flux, since the behaviour of
the system is the same for other values by simply shifting the
temperature range \cite{gosalvez2017}. The simulated surfaces contain
$283 \times 283$ Cartesian units, corresponding to
$L_{x} \times L_{y} = 283 \times 326 = 92258$ adsorption sites in the
triangular lattice, and periodic boundary conditions are applied. The
simulations are evolved until 100 \% coverage ($\theta=$ 1 ML),
repeating them $K=10$ times in order to obtain ensemble averages for
all quantities of interest. Strictly two-dimensional growth is
simulated (no three-dimensional features are attempted). Snapshots of
the surface configuration are obtained every 5 \% of coverage, which
are used as input for the morphology analysis (see the end of section
\ref{sec:introduction}).


\section{Results}

\subsection{Island density}
\label{sec:IslandDensity}

\begin{figure}[htb!]
  \def\svgwidth{\columnwidth}
\executeiffilenewer{IslandsVsFluxBoth_36_0_100000.svg}{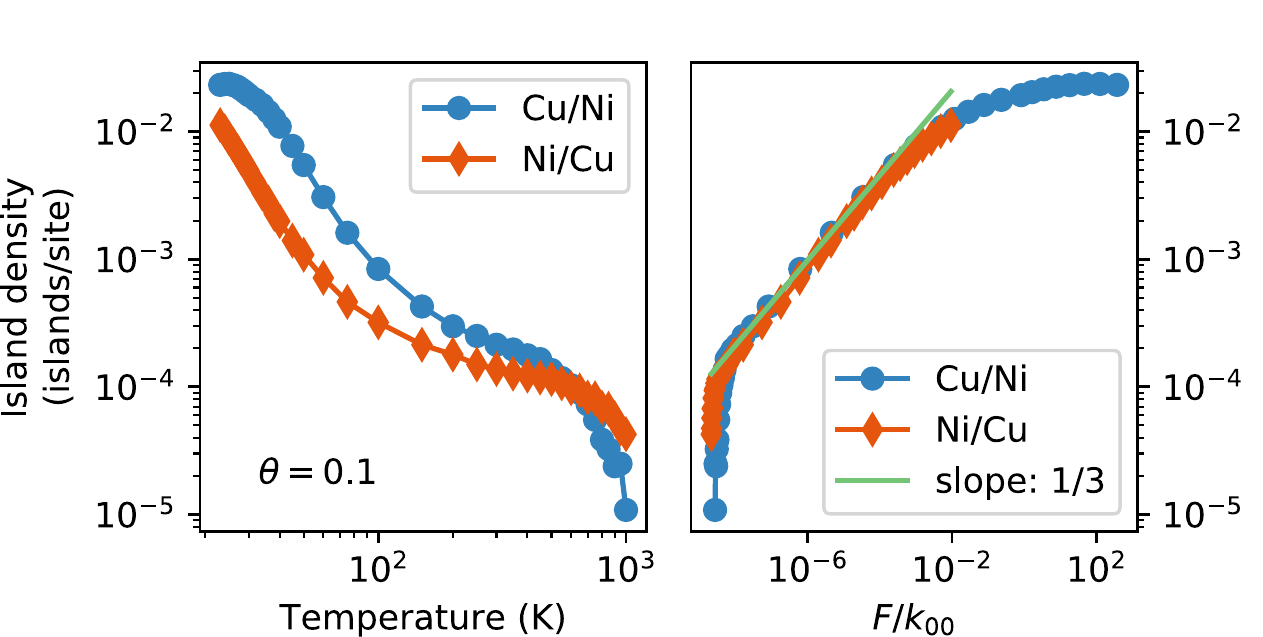}%
{inkscape             -z             -D             --file=IslandsVsFluxBoth_36_0_100000.svg             %
--export-pdf=IslandsVsFluxBoth_36_0_100000.pdf             --export-latex}%
\begingroup%
  \makeatletter%
  \providecommand\color[2][]{%
    \errmessage{(Inkscape) Color is used for the text in Inkscape, but the package 'color.sty' is not loaded}%
    \renewcommand\color[2][]{}%
  }%
  \providecommand\transparent[1]{%
    \errmessage{(Inkscape) Transparency is used (non-zero) for the text in Inkscape, but the package 'transparent.sty' is not loaded}%
    \renewcommand\transparent[1]{}%
  }%
  \providecommand\rotatebox[2]{#2}%
  \newcommand*\fsize{\dimexpr\f@size pt\relax}%
  \newcommand*\lineheight[1]{\fontsize{\fsize}{#1\fsize}\selectfont}%
  \ifx\svgwidth\undefined%
    \setlength{\unitlength}{363.5693335bp}%
    \ifx\svgscale\undefined%
      \relax%
    \else%
      \setlength{\unitlength}{\unitlength * \real{\svgscale}}%
    \fi%
  \else%
    \setlength{\unitlength}{\svgwidth}%
  \fi%
  \global\let\svgwidth\undefined%
  \global\let\svgscale\undefined%
  \makeatother%
  \begin{picture}(1,0.50432264)%
    \lineheight{1}%
    \setlength\tabcolsep{0pt}%
    \put(0,0){\includegraphics[width=\unitlength,page=1]{IslandsVsFluxBoth_36_0_100000.pdf}}%
    \put(0.16063982,0.46241824){\color[rgb]{0,0,0}\makebox(0,0)[lt]{\lineheight{1.25}\smash{\begin{tabular}[t]{l}\begin{subfigure}{0.0\columnwidth}    \caption{~}    \label{fig:cuniIslands}  \end{subfigure}  \end{tabular}}}}%
    \put(0.54964007,0.46241824){\color[rgb]{0,0,0}\makebox(0,0)[lt]{\lineheight{1.25}\smash{\begin{tabular}[t]{l}\begin{subfigure}{0.0\columnwidth}    \caption{~}    \label{fig:cuniIslandsFlux}  \end{subfigure}  \end{tabular}}}}%
  \end{picture}%
\endgroup%

  \caption{Average island density at 10 \% coverage for Cu/Ni(111) and
    Ni/Cu(111) from 23 K to 1000 K with concerted island and
    multi-atom diffusion events activated.}
  \label{fig:cunicuIslandsBoth}
\end{figure}

We first consider the island density, $\langle n_{\rm{isl}} \rangle$,
defined for any given coverage as the ensemble average of the total
number of islands divided by the total number of adsorption sites
$L_{x}L_{y}$.  Figure \ref{fig:cuniIslands} shows that
$\langle n_{\rm{isl}} \rangle$ is higher for Cu/Ni(111) than for
Ni/Cu(111) in all the temperature range, except for the highest
temperatures. The plot corresponds to 10 \% coverage
($\theta = 0.10$), which is low enough to avoid potential coalescence
of neighbour islands, while it is high enough to ensure the formation
of stable islands. For both systems, the lower the temperature the
larger the island density and, overall, the temperature dependence is
similar. Nevertheless, for the same coverage and temperature, the
particular value of the density is different.

This behaviour agrees well with traditional 2D nucleation theory
\cite{venables_nucleation_1987, antczak_surface_2010}:
\begin{equation}
  \label{eq:numberOfIslands}
  n_{\rm{isl}} \propto \left(\frac{F}{D}\right)^\frac{i}{i+2} , 
\end{equation}
where $F$ is the adsorption flux,
$D=\frac{1}{2 \delta} \frac{\langle \mathcal{R}^2 \rangle}{t}$ is the
diffusivity for a single monomer ($D= \frac{3}{4} k_m l^2$ for the
triangular lattice, with $k_m$ the monomer diffusion rate and $l = 1$
the hop distance) and $i$ is the critical island size (islands with
size $>i$ are stable). In our case, the deposition flux $F$ takes the
same value and the critical island size $i$ behaves practically the
same in both systems (see next paragraph). However, $D \propto k_{m}$
varies with the adatom type (Ni or Cu) and temperature. In fact, the
monomer diffusion energy barrier is 52 meV for Cu/Ni(111) and 31 meV
for Ni/Cu(111), which implies higher $D$ for the Ni/Cu(111) system
and, thus, lower island density. Therefore, the observed behaviour
with temperature agrees with expectations, except at the highest
temperatures, where diffusion is not controlled anymore by the
monomers and the critical island size deviates from one system to the
other.

In fact, the plot of $\log( n_{\rm{isl}} )$ {\it vs} $\log(F/k_{m})$
in figure \ref{fig:cuniIslandsFlux} shows that the two systems follow
equation \ref{eq:numberOfIslands}, with $i=1$ and
$x=\frac{i}{1+2}=\frac{1}{3}$ at medium temperatures (= medium
$F/k_{m}$). This means that dimers are the smallest stable nuclei in
this range. At low temperatures (high $F/k_{m}$), $n_{\rm{isl,Ni}}$
displays a tendency towards saturation, indicating that $i \approx 0$
for extremely low temperatures, i.e.\ monomers already form stable
nuclei. This is due, literally, to the absence of diffusion and the
dominant role of adsorption, as will be shown in section
\ref{sec:Results_R}. Note that $n_{\rm{isl,Cu}}$ shows the same
tendency at low temperatures. In turn, at high temperatures (low
$F/k_{m}$), the slopes of $n_{\rm{isl,Ni}}$ and $n_{\rm{isl,Cu}}$
increase dramatically while slightly deviating from each other,
indicating, as expected, that significantly more than two adatoms are
required to stabilise a cluster and the actual diffusion events
contributing to the stabilisation of the nuclei differ slightly from
one system to the other.

\subsection{Morphology}

Not only the island density differs from one system to the other,
their morphology deviates as well. This is shown in figures
\ref{fig:cuni} and \ref{fig:nicu} for a collection of
representative temperatures at $\theta = 0.10$. At the lower
temperatures the islands are more dendritic in both systems,
reflecting low diffusivity along the island perimeters after monomer
attachment. At the higher temperatures, however, the islands tend to
be compact/hexagonal, reflecting high diffusivity at the
perimeters. In this case, the adatoms move quickly along the
perimeters and are able to find the lowest energy sites (or
thermodynamically stable positions).

The morphology of the islands reflects differences in the growth
process. Simple visual inspection indicates that the island shapes are
different, specially at 350 K, where Cu forms compact islands while Ni
condensates into dendritic shapes. At the other temperatures, however,
the distinction is less obvious and a quantitative PSD analysis is
required to show the actual variations. By using the images from
$K=10$ equivalent simulations with different random numbers, the
corresponding PSD maps are shown in figure \ref{fig:cuniPsd} for
Cu/Ni(111) and figure \ref{fig:nicuPsd} for Ni/Cu(111). In addition,
point-by-point PSD difference maps are shown in figure
\ref{fig:diffCuNiCuPsd}.

\begin{figure}[htb!]
  {\small
    \def\svgwidth{1.1\columnwidth}
\executeiffilenewer{surfaces.svg}{surfaces.pdf}%
{inkscape             -z             -D             --file=surfaces.svg             %
--export-pdf=surfaces.pdf             --export-latex}%
\input{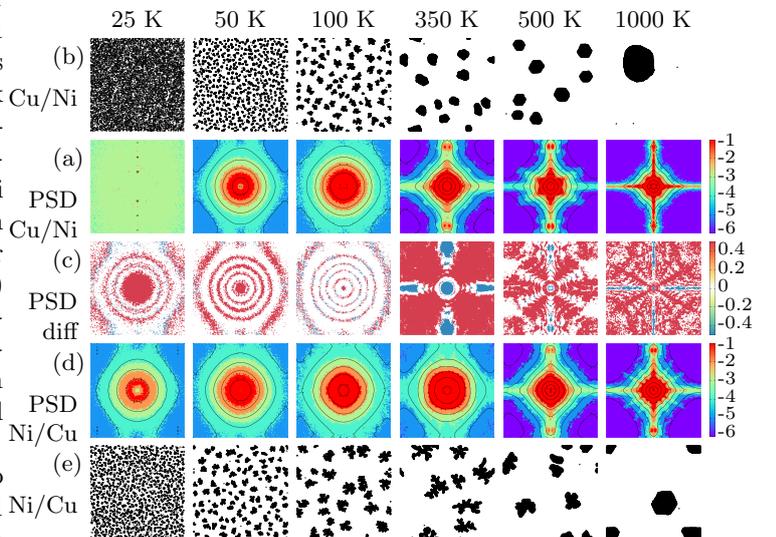}%

  }
  \caption{(\subref{fig:cuni}),(\subref{fig:nicu}) Surface morphology
    for Cu/Ni(111) and Ni/Cu(111), respectively, at $\theta = 0.10$
    and various temperatures (as indicated) with concerted island and
    multi-atom diffusion events
    activated. (\subref{fig:cuniPsd}),(\subref{fig:nicuPsd}) PSD maps
    for Cu/Ni(111) (\subref{fig:cuniPsd}) and Ni/Cu(111)
    (\subref{fig:nicuPsd}). (\subref{fig:diffCuNiCuPsd}) PSD
    difference maps.}
  \label{fig:cunicu}
\end{figure}

At 25 K, where visual inspection is difficult, the PSD difference map
displays several circular patterns, clearly deviating from random
fluctuations around the 0 value (noise) and, thus, concluding that the
two surfaces differ structurally. Note that perfect noise on the PSD
difference map is indicated by random blue/red/white values associated
with positive/negative/zero fluctuations between the two maps. At 50
K, the PSD difference map is essentially the same as for 25 K, thus
revealing structural differences. At 100 K, where both PSD maps are
most similar, the difference map still shows circles. At the already
considered temperature of 350 K, the Cu/Ni(111) map displays
considerably higher values, except at the central and cross-like
regions, where it is lower. At 500 K, the PSD difference map still
reveals a strong structural mismatch. Here, the Cu/Ni(111) islands are
almost hexagonal while the Ni/Cu(111) islands still remain
amorphous. At the highest considered temperature (1000 K), both
islands are compact. However, the shape for Cu/Ni(111) resembles a
circle while that for Ni/Cu(111) approaches a hexagon; the PSD map for
Cu/Ni(111) is mostly higher than that for Ni/Cu(111), with significantly 
lower values at the center and at four elongated horizontal/vertical regions. 
Overall, comparing
the two systems at the same temperature and coverage, we conclude that
they display different island densities and morphologies.
\subsection{Total rate}
\label{sec:Results_R}

\begin{figure}[htb!]
  \centering
  \begin{subfigure}{0.9\columnwidth}
    \caption{\hspace{10cm} }
    \def\svgwidth{\textwidth}
\executeiffilenewer{TotalRates10both0_1.svg}{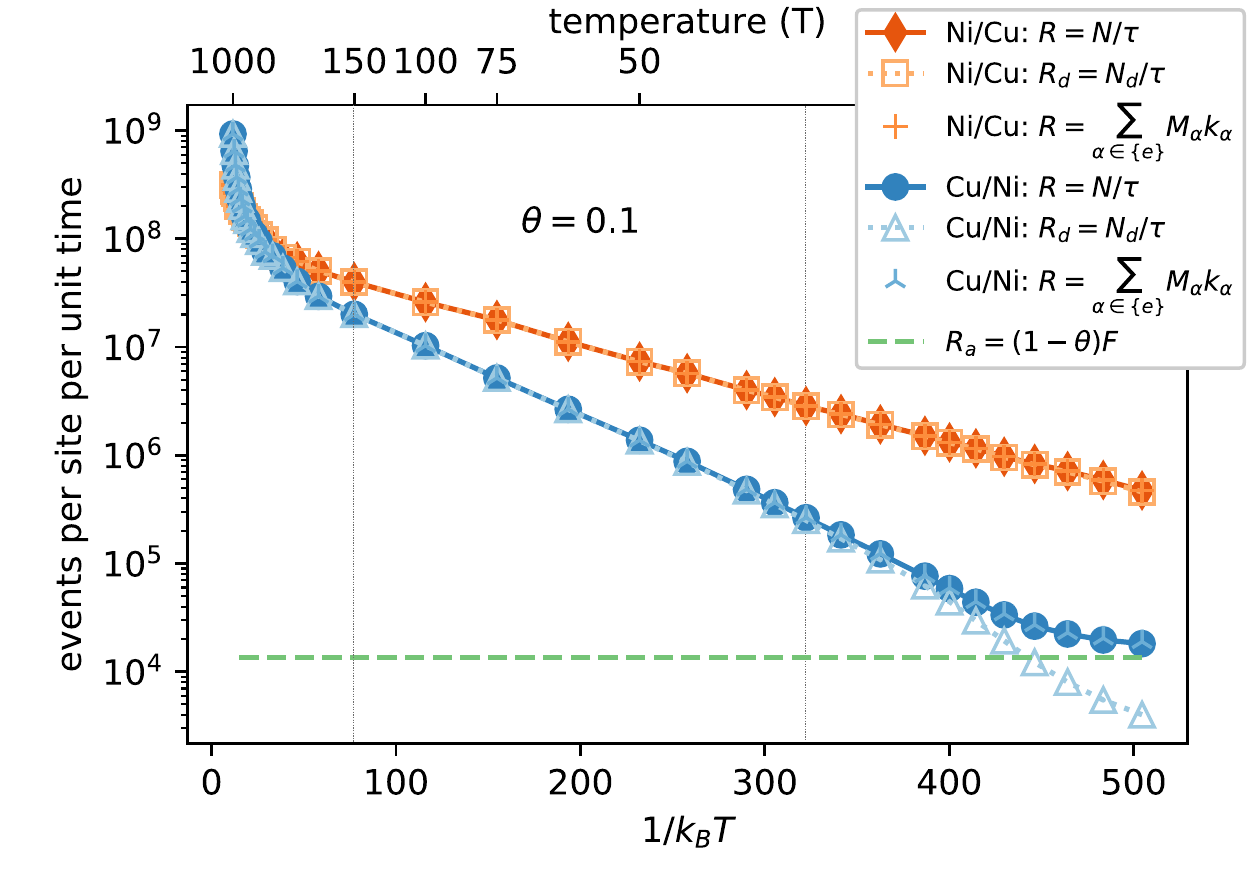}%
{inkscape             -z             -D             --file=TotalRates10both0_1.svg             %
--export-pdf=TotalRates10both0_1.pdf             --export-latex}%
\begingroup%
  \makeatletter%
  \providecommand\color[2][]{%
    \errmessage{(Inkscape) Color is used for the text in Inkscape, but the package 'color.sty' is not loaded}%
    \renewcommand\color[2][]{}%
  }%
  \providecommand\transparent[1]{%
    \errmessage{(Inkscape) Transparency is used (non-zero) for the text in Inkscape, but the package 'transparent.sty' is not loaded}%
    \renewcommand\transparent[1]{}%
  }%
  \providecommand\rotatebox[2]{#2}%
  \newcommand*\fsize{\dimexpr\f@size pt\relax}%
  \newcommand*\lineheight[1]{\fontsize{\fsize}{#1\fsize}\selectfont}%
  \ifx\svgwidth\undefined%
    \setlength{\unitlength}{360bp}%
    \ifx\svgscale\undefined%
      \relax%
    \else%
      \setlength{\unitlength}{\unitlength * \real{\svgscale}}%
    \fi%
  \else%
    \setlength{\unitlength}{\svgwidth}%
  \fi%
  \global\let\svgwidth\undefined%
  \global\let\svgscale\undefined%
  \makeatother%
  \begin{picture}(1,0.7)%
    \lineheight{1}%
    \setlength\tabcolsep{0pt}%
    \put(0,0){\includegraphics[width=\unitlength,page=1]{TotalRates10both0_1.pdf}}%
    \put(0.20323476,0.11639286){\color[rgb]{0.42352941,0.42352941,0.42352941}\makebox(0,0)[lt]{\lineheight{1.25}\smash{\begin{tabular}[t]{l}A\end{tabular}}}}%
    \put(0.44929911,0.11639286){\color[rgb]{0.42352941,0.42352941,0.42352941}\makebox(0,0)[lt]{\lineheight{1.25}\smash{\begin{tabular}[t]{l}B\end{tabular}}}}%
    \put(0.78341853,0.11680951){\color[rgb]{0.42352941,0.42352941,0.42352941}\makebox(0,0)[lt]{\lineheight{1.25}\smash{\begin{tabular}[t]{l}C\end{tabular}}}}%
    \put(0,0){\includegraphics[width=\unitlength,page=2]{TotalRates10both0_1.pdf}}%
  \end{picture}%
\endgroup%

    \label{fig:nicutotalrates10}
  \end{subfigure}
  \begin{subfigure}{0.9\columnwidth}
    \caption{\hspace{10cm} }
    \def\svgwidth{\textwidth}
\executeiffilenewer{TotalRates10both0_09.svg}{TotalRates10both0_09.pdf}%
{inkscape             -z             -D             --file=TotalRates10both0_09.svg             %
--export-pdf=TotalRates10both0_09.pdf             --export-latex}%
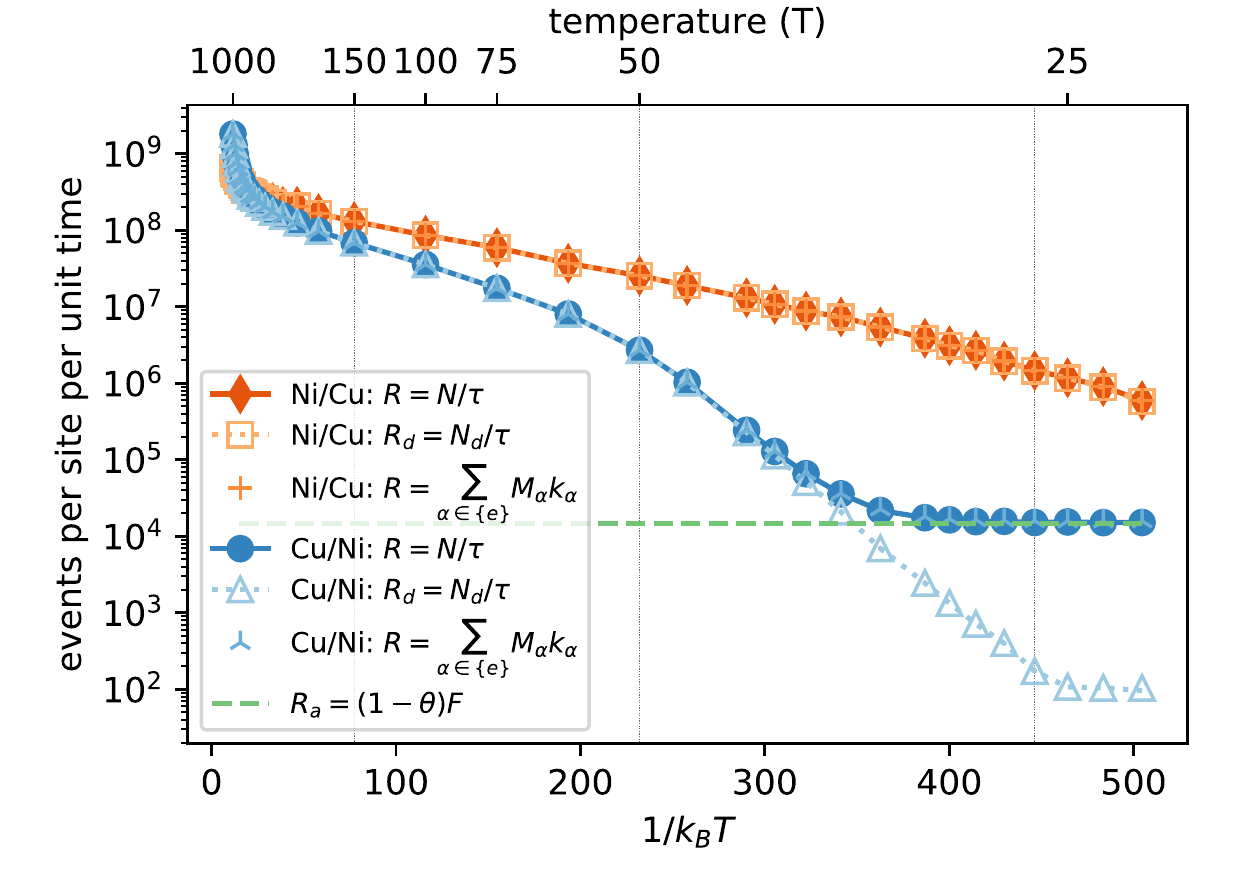%

    \label{fig:nicutotalrates01}
  \end{subfigure}
  \caption{Average total rates per site for diffusion ($R_{h}$),
    adsorption ($R_{a}$) and all events ($R = R_{h} + R_{a}$) for
    Cu/Ni(111) and Ni/Cu(111), as indicated,
    (\subref{fig:nicutotalrates10}) at 10 \% coverage and
    (\subref{fig:nicutotalrates01}) at 1 \% coverage.}
  \label{fig:nicutotalrates}
\end{figure}

In addition to the differences in the island density and morphology,
also the average total rate per site, $R = N / \tau$ (equation
\ref{eq:AverageRe1}), differs between the two systems. This is shown
in the Arrhenius plot of figure \ref{fig:nicutotalrates10} for
$\theta = 0.10$ and $T = 23 - 1000$ K, while the case for
$\theta = 0.01$ is shown in figure \ref{fig:nicutotalrates01} and many
other coverage values are considered in figures
\ref{fig:allTotalRates} and \ref{fig:allTotalRatesMap} of the
appendix. These figures also display the average total rate per site
determined using equation \ref{eq:AverageRe2},
$R = \sum_{\alpha \in \{ e \} } M_{\alpha} k_{\alpha}$, demonstrating
that both equations \ref{eq:AverageRe1} and \ref{eq:AverageRe2}
provide equivalent descriptions of the same quantity. In addition, the
figures also show the average total diffusion rate per site,
$R_{d} = N_{d} / \tau$
($= \sum_{\alpha \in \{ h \} } M_{\alpha} k_{\alpha}$) for the two
systems and the average total adsorption rate per site,
$R_{a} = N_{a} / \tau $, which is identical for both systems and
independent of temperature, only depending on coverage:
$R_{a} = M_{a} k_{a} = (1-\theta) F$.

Regarding figure \ref{fig:nicutotalrates10}, the total rate is much
higher in the Ni/Cu(111) system, specially at low temperatures (e.g.\
region C). Since adsorption is identical in both systems and remains
quite low, the difference in their total rate is primarily due to the
total diffusion rate, which is higher for Ni/Cu(111). Nevertheless, in
the Cu/Ni(111) system adsorption plays an important role at the lowest
temperatures (region C), where it provides the largest contribution to
the total rate, significantly over the total diffusion rate. In fact,
the adsorption rate ($1.5 \times 10^4$ Hz) is higher than the monomer
diffusion rate at 25 K ($3.3 \times 10^2$ Hz for a diffusion barrier
of 52 meV). This behaviour is noticeable until about 32 K (the
frontier between regions B and C), above which the total rate is
essentially dominated by the total diffusion rate, as for the
Ni/Cu(111) over the whole considered range of temperature.

At lower coverage, more complex behaviour is observed at low
temperatures, as shown in regions C and D of figure
\ref{fig:nicutotalrates01}, especially in the case of $R_{d}$ for
Cu/Ni(111). As shown in section \ref{sec:Results_ae} below, the
diffusivity in this case ($R_{d}$ for Cu/Ni(111)) is dominated by
non-concerted dimer diffusion in regions C and D and it is ruled by
monomer diffusion in region B. In region A, monomer diffusion is
complemented by perimeter diffusion and both concerted and
non-concerted dimer diffusion, in addition to other secondary
events. Although non-concerted dimer diffusion dominates in both
regions C and D, it has not yet really been activated in region D.
The behaviour for the total rate $R$ of Cu/Ni(111) is similar to that
of $R_{d}$, but $R$ remains higher than $R_{d}$ at low temperatures
due to the larger value of the total adsorption rate
($R_{a}$). Finally, the trend for Ni/Cu(111) in figure
\ref{fig:nicutotalrates01} is similar, but displaced towards lower
temperatures.
\subsection{Activation Energy}
\label{sec:Results_ae}

\begin{figure}[htb!]
  \centering  
  \begin{subfigure}{0.99\columnwidth}
    \caption{\hspace{10cm} }
    \label{fig:aenicu10}
    \def\svgwidth{\textwidth}
\executeiffilenewer{niCuMultiplicitiesResumeT0_1.svg}{niCuMultiplicitiesResumeT0_1.pdf}%
{inkscape             -z             -D             --file=niCuMultiplicitiesResumeT0_1.svg             %
--export-pdf=niCuMultiplicitiesResumeT0_1.pdf             --export-latex}%
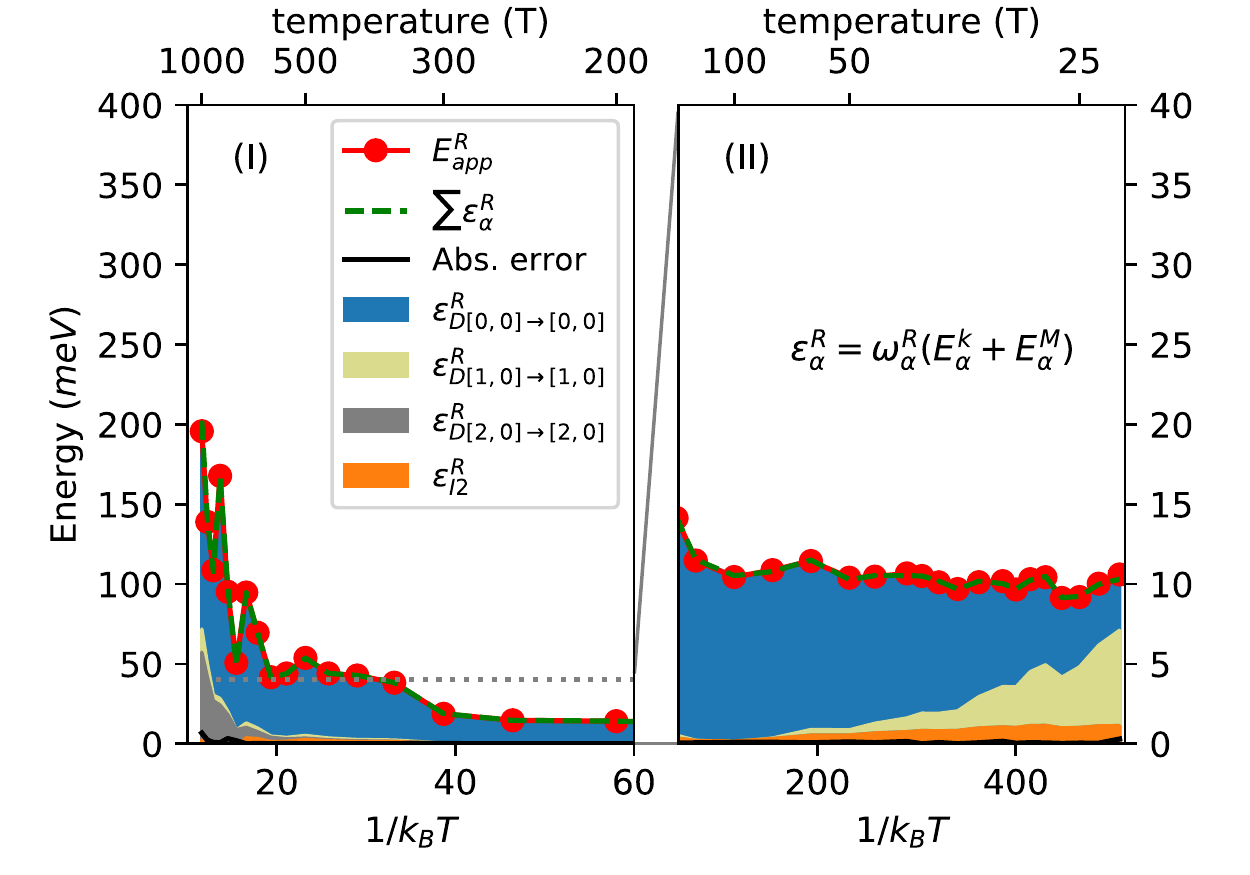%
    
  \end{subfigure}
  \begin{subfigure}{0.99\columnwidth}
    \caption{\hspace{10cm} }
    \label{fig:aecuni10}
    \def\svgwidth{\textwidth}
\executeiffilenewer{cuNiMultiplicitiesResumeT0_1.svg}{cuNiMultiplicitiesResumeT0_1.pdf}%
{inkscape             -z             -D             --file=cuNiMultiplicitiesResumeT0_1.svg             %
--export-pdf=cuNiMultiplicitiesResumeT0_1.pdf             --export-latex}%
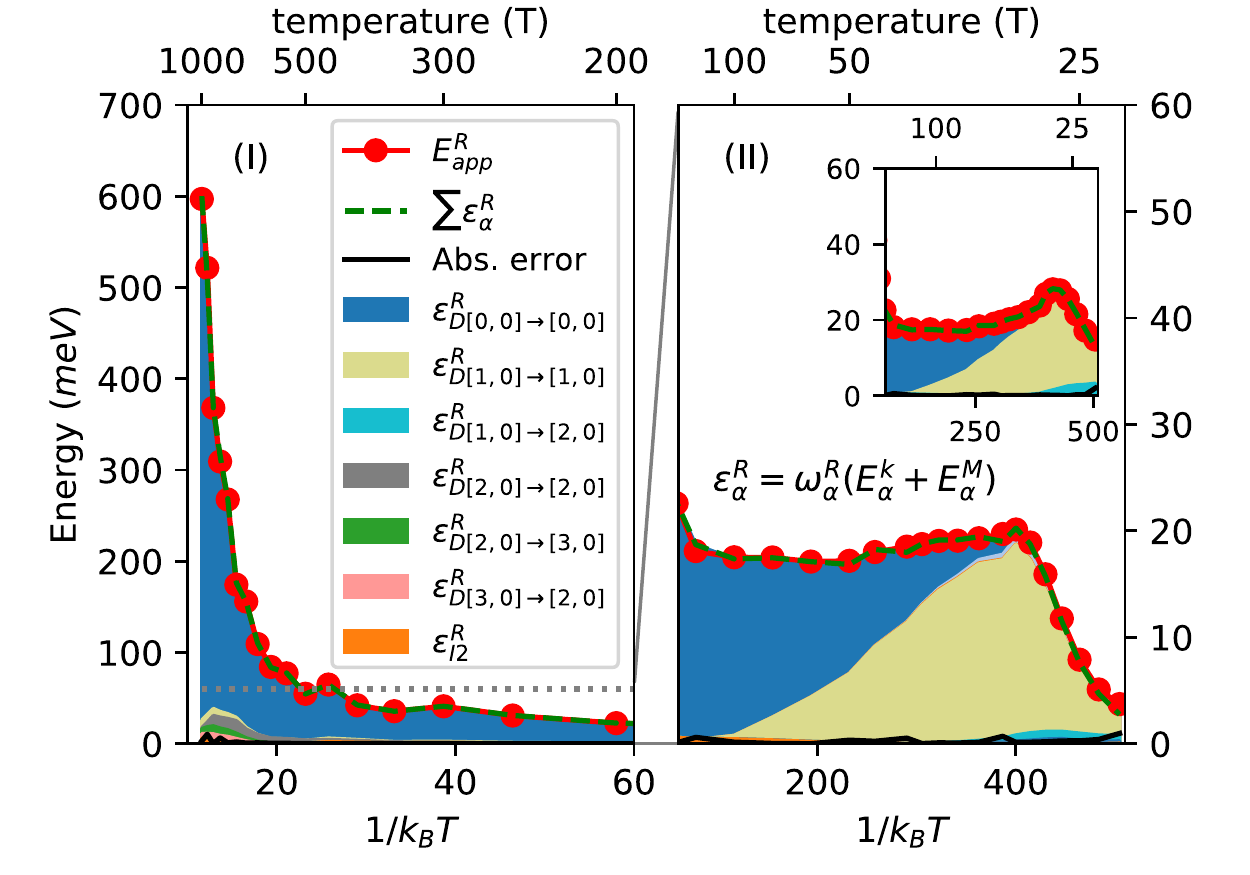%
    
  \end{subfigure} 
  \caption{Temperature dependence of the apparent activation energy of
    the average total rate per site ($E_{app}^R$) at $\theta = 0.10$
    for (\subref{fig:aenicu10}) Ni/Cu(111) and (\subref{fig:aecuni10})
    Cu/Ni(111). $E_{app}^R$ is described well by
    $\sum_{\alpha \in \{ e \}} \epsilon_\alpha^R$, where
    $\epsilon_\alpha^R = \omega_\alpha^R (E_\alpha^{k} +
    E_\alpha^{M})$. The absolute error
    $| E_{app}^R - \sum_{\alpha \in \{ e \}} \epsilon_\alpha^R
    |$ is also plotted.}
  \label{fig:ae10}
\end{figure}

For the Arrhenius plot in figure \ref{fig:nicutotalrates10}, the slope
of $R$ {\it vs} $\beta$ is the apparent activation energy,
$E_{app}^R$, which is shown in figure \ref{fig:aenicu10} for
Ni/Cu(111) and figure \ref{fig:aecuni10} for Cu/Ni(111). While these
plots correspond to $\theta = 0.10$ and $T = 23 - 1000$ K, similar
results for additional coverage values are shown in figures
\ref{fig:aeStudyCuNi} and \ref{fig:aeStudyNiCu} of the appendix. In
each plot, we show two temperature regions: (I) $1000 \ge T > 150$ K,
and (II) $150 \ge T \ge 23$ K, with the low temperature region
displayed in a magnified view. In addition, each region shows two
alternative expressions for the apparent activation energy, namely,
$E_{app}^R = - \frac{ \partial \log R} {\partial \beta}$ with
$R = N / \tau$ (equations \ref{EappR_0} and \ref{eq:AverageRe1}) and
$E_{app}^{R,*} = \sum_{\alpha \in \{ e\}} \epsilon_\alpha^R$ with
$\epsilon_\alpha^R = \omega_\alpha^R(E_\alpha^k + E_\alpha^M)$
(equation \ref{EappR_3}). The former ($E_{app}^R$ in the plots) is
obtained numerically by using finite central differences of $\log R$
and $\beta$. The latter ($\sum_{\alpha \in \{ e\}} \epsilon_\alpha^R$
in the plots) is obtained from the rate of every event type,
$k_\alpha$, and the corresponding multiplicity, $M_\alpha$, in order
to determine the probability of every event type, $\omega_\alpha^R$
(equation \ref{omega_definition}), as well as by summing the energy
barrier, $E_\alpha^{k}$, and the configurational contribution,
$E_\alpha^{M} = -\frac{ \partial \log M_\alpha }{ \partial \beta }$,
calculated by finite differences as well. In addition, each plot shows
the absolute error between the two measures,
$| E_{app}^R - \sum_{\alpha \in \{ e\}} \epsilon_\alpha^R |$, which
remains smaller than 6.65 meV for Ni/Cu(111) and 3.12 meV for
Cu/Ni(111), with a mean value of 0.49 meV for Ni/Cu(111) and 0.51 meV
for Cu/Ni(111). The maximum error is typically due to the finite
difference estimate of the slope (not the multiplicity based formula)
and it usually occurs at the highest/lowest temperature or when
$\log R$ fluctuates with respect to the previous temperature. Thus,
figures \ref{fig:aenicu10} and \ref{fig:aecuni10} show that equation
\ref{EappR_3} accurately explains the values observed for the apparent
activation energy.

Next, we analyse the different contributions to the apparent activation 
energy. Before that, however, it is useful to note that,
for Ni/Cu(111) in figure \ref{fig:aenicu10}, the apparent activation
energy of the total rate, $E_{app}^R$, is also the apparent
activation energy of the total diffusion rate,
$E_{app}^{R_d} = E_{app}^R$, since $R = R_d$ in this system (see
figure \ref{fig:nicutotalrates10}). In turn, based on equation
\ref{EappR_d_0}, the apparent activation energy of the diffusivity is:
$E_{app}^{D_{T}} \approx E_{app}^{R_d} = E_{app}^R$, since the
contribution from the correlation factor,
$E^{f_{T}} = -\frac{ \partial \log f_{T} }{ \partial \beta }$, is very
small \cite{gosalvez2017}.

\begin{figure}[htb!]
  \centering
  \begin{subfigure}{0.9\columnwidth}
    \caption{\hspace{10cm} }
    \label{fig:omeganicu10}
    \def\svgwidth{\textwidth}
\executeiffilenewer{nicuPlot36_0_1T1.svg}{nicuPlot36_0_1T1.pdf}%
{inkscape             -z             -D             --file=nicuPlot36_0_1T1.svg             %
--export-pdf=nicuPlot36_0_1T1.pdf             --export-latex}%
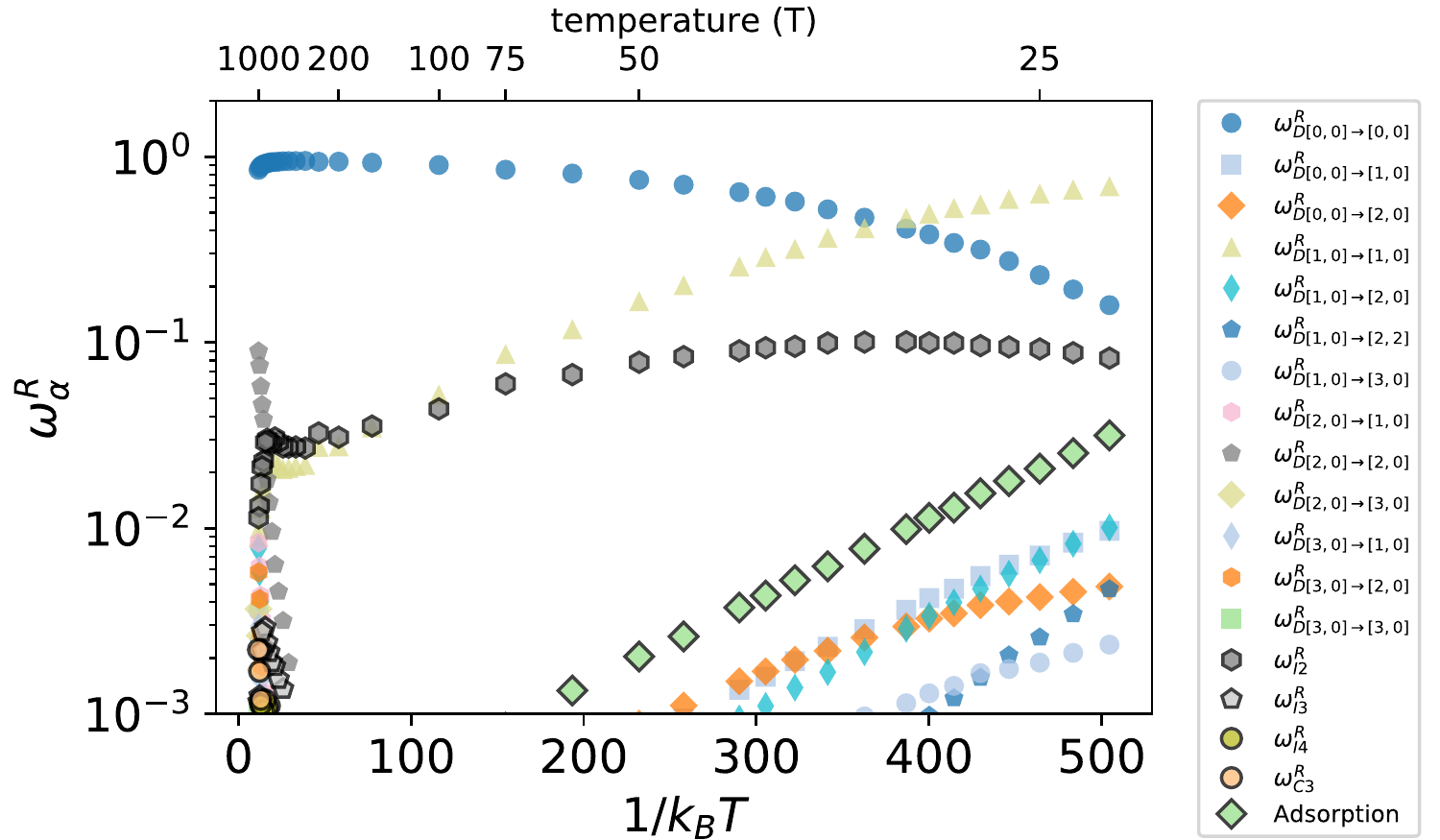%
    
  \end{subfigure}
  \begin{subfigure}{0.90\columnwidth}
    \caption{\hspace{10cm} }
    \label{fig:omegacuni10}
    \def\svgwidth{\textwidth}
\executeiffilenewer{cuniPlot36_0_1T1.svg}{cuniPlot36_0_1T1.pdf}%
{inkscape             -z             -D             --file=cuniPlot36_0_1T1.svg             %
--export-pdf=cuniPlot36_0_1T1.pdf             --export-latex}%
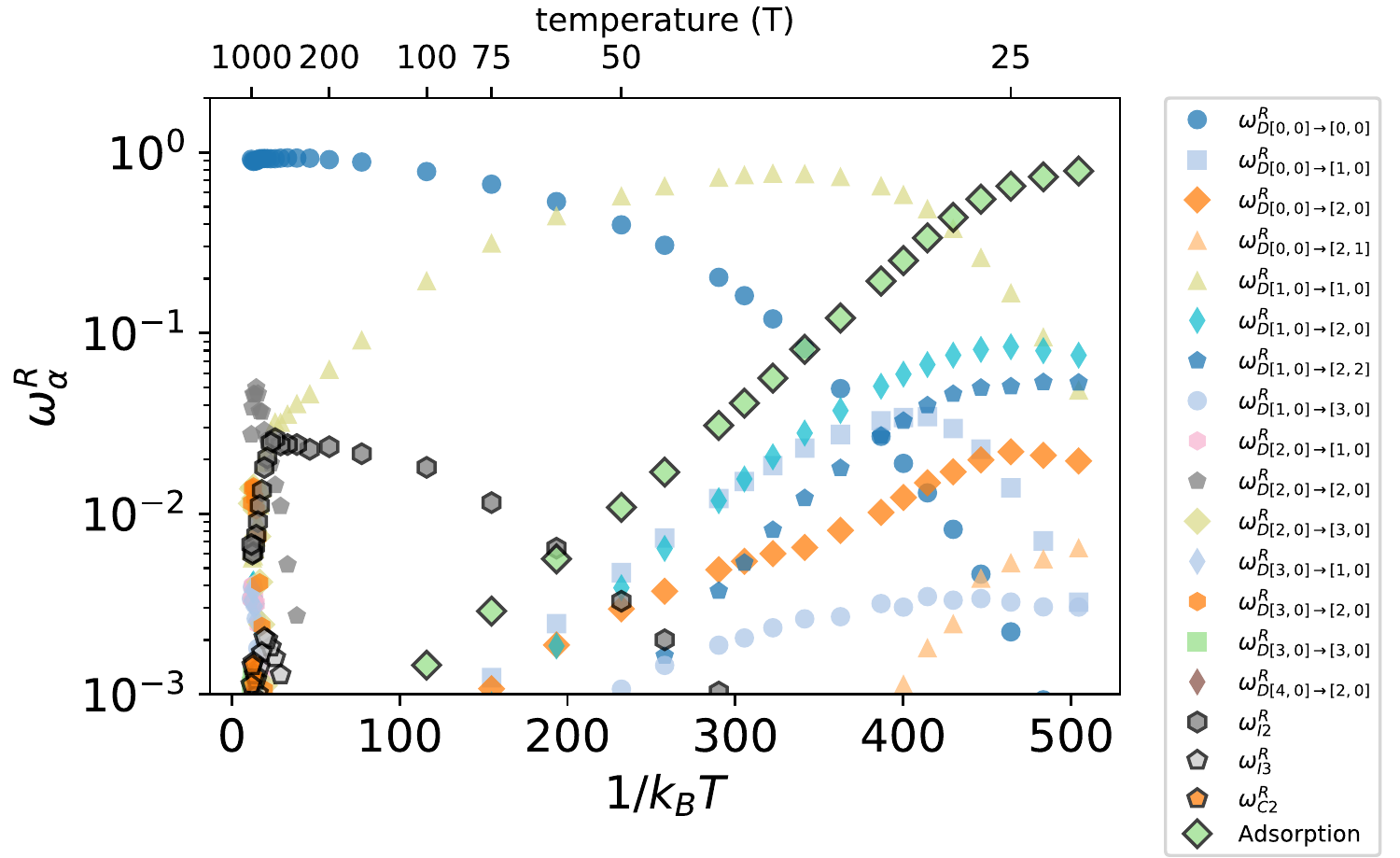%
    
  \end{subfigure}
  \caption{Temperature dependence of the event probabilities,
    $w_\alpha^R$, at $\theta = 0.10$ for
    (\subref{fig:omeganicu10}) Ni/Cu(111) and
    (\subref{fig:omegacuni10}) Cu/Ni(111). Only those events whose
    probability is higher than $10^{-3}$ are shown.}
  \label{fig:omegaLog10}
\end{figure}

Since the apparent activation energy is constant ($\approx 10$ meV) in
region II of figure \ref{fig:aenicu10} for Ni/Cu(111), traditionally
one would be tempted to conclude that there is a single
rate-controlling event in this temperature range. However, 10 meV does
not correspond to any of the energy barriers included in the
system. In fact, the multiplicity analysis based on equation
\ref{EappR_3},
$E_{app}^R = \sum_{\alpha \in \{ e \}} \epsilon_\alpha^R =
\sum_{\alpha \in \{ e \}} \omega_\alpha^R(E_\alpha^k +
E_\alpha^M)$, shows that there are three main contributing events in
this region, namely, monomer diffusion ($D[0,0] \rightarrow [0,0]$,
with $E_{D[0,0] \rightarrow [0,0]}^{k} = 31$ meV), non-concerted dimer
diffusion ($D[1,0] \rightarrow [1,0]$, with
$E_{D[1,0] \rightarrow [1,0]}^{k} =16$ meV), and concerted dimer
diffusion ($I2$, with $E_{I2}^{k} =21$ meV). The major contribution
shifts from (non-concerted + concerted) dimer diffusion at the lowest
temperatures (where the chance to form dimers is high) towards monomer
diffusion at the highest temperatures in this range (where recently
adsorbed monomers have a larger chance to reach an island than to form
a dimer). Note that the shift is mostly due to the change in the event
probabilities, $\omega_\alpha^R$, with temperature for those three
particular event types, as shown in figure \ref{fig:omeganicu10}. In
addition, the configurational contributions to the apparent energy for
the three event types, $E_{D[0,0] \rightarrow [0,0]}^{M}$,
$E_{D[1,0] \rightarrow [1,0]}^{M}$ and $E_{I2}^{M}$, are negative in
this case, thus leading to a value of the apparent activation energy
($\approx 10$ meV) that is significantly smaller than any of the three
energy barriers (31, 16 and 21 meV).

For the lowest temperatures ($T < 60$ K), figure \ref{fig:omeganicu10}
shows that several additional event types have appreciable roles, with
probabilities larger than 0.1\% and up to about 3\%. This includes
adsorption (no energy barrier), monomer attachment to the islands
($D[0,0] \rightarrow [1,0]$ and $D[0,0] \rightarrow [2,0]$, with
energy barriers of 28 and 15 meV, respectively) and perimeter adatom
stabilization ($D[1,0] \rightarrow [2,0]$, 14 meV;
$D[1,0] \rightarrow [2,2]$, 0 meV; and $D[1,0] \rightarrow [3,0]$, 1
meV). Furthermore, concerted dimer diffusion ($I2$, with
$E_{I2}^{k} =21$ meV) has an appreciable role over the complete
temperature range, with an event probability of 8-10\% up to about 50
K, and remaining active at higher temperatures ($\ge$ 1\%). Finally,
at the highest temperatures, edge diffusion
($D[2,0] \rightarrow [2,0]$, with
$E_{D[2,0] \rightarrow [2,0]}^{k} =364$ meV), trimer diffusion ($I3$,
with $E_{I3}^{k}=148$ meV) and a few other processes become relevant,
with event probabilities larger than 0.1\%.

Here, the energy barrier for non-concerted dimer diffusion (16 meV) is
smaller than that for concerted dimer diffusion (21 meV) and, thus,
non-concerted diffusion has a larger rate, especially at low
temperatures (e.g.\ $k_{D[1,0] \rightarrow [1,0]} = 3.1 \times 10^9$
Hz and $k_{I2} = 2.5 \times 10^8$ Hz at 23 K). On the other hand, the
multiplicities of the two event types are similar, with
$M_{D[1,0] \rightarrow [1,0]} = 4n_{2}$ (where $n_{2}$ is the density
of dimers and both atoms may hop in two directions while remaining
attached to the other, thus leading to a multiplicity of
$2 \cdot 2 = 4$ per dimer) while $M_{I2} = 6n_{2}$ (since there are
six hop directions in the triangular lattice). Thus, comparing the
total rates per site for both events,
$4n_{2}k_{D[1,0] \rightarrow [1,0]}$ and $6n_{2}k_{I2}$, non-concerted
dimer diffusion occurs more often at low temperatures. However, at
high temperatures the two rates become very similar and concerted
dimer diffusion occurs more frequently due to the slightly larger
multiplicity. See figure \ref{fig:dimers} in the appendix for further
proof. Based on this example, we believe that there may be systems
where concerted dimer diffusion dominates over non-concerted diffusion
in a wide range of temperature.

For Cu/Ni(111) in figure \ref{fig:aecuni10}, the situation is very
similar, except for the fact that $E_{app}^R$ approaches zero at the
low temperature end. In this region, the total rate is dominated by
adsorption, $R = R_{a} = M_{a}k_{a}$ (see figure
\ref{fig:nicutotalrates10}), with both the adsorption rate,
$k_{a} = F$, and the multiplicity, $M_{a}=1-\theta$, being temperature
independent. Thus,
$\epsilon_{a}^R = \omega_{a}^R(E_{a}^k + E_{a}^M) = 0$, because
$E_{a}^k = E_{a}^M = 0$ even though $w_{a}^R \approx 1$ (see figure
\ref{fig:omegacuni10}).

The insert in figure \ref{fig:aecuni10} for $\theta = 0.1$ displays
the apparent activation energy of the total diffusion rate per site,
$E_{app}^{R_{d}}$, characterised by a weak maximum in the range 23-50
K, which is clearly assigned to the temperature dependence of
$\epsilon_{D[1,0] \rightarrow [1,0]}^R$ (non-concerted dimer
diffusion). Note that a similar maximum is observed in the insert of
figure \ref{fig:aeStudyCuNi} for $\theta = 0.01$ in the appendix and
it is also assigned to $\epsilon_{D[1,0] \rightarrow [1,0]}^R$.  For
such low temperatures, the quickly-diffusing dimers (through
non-concerted diffusion) collide against the slowly-diffusing monomers
(which essentially act as stationary obstacles). This generates both
triangular (immobile) and chain-like (mobile) trimers, which are
eventually compacted into the triangular (immobile) shape {\it via}
perimeter adatom stabilization ($D[1,0] \rightarrow [2,0]$ and
$D[1,0] \rightarrow [2,2]$). The maxima in
$\epsilon_{D[1,0] \rightarrow [1,0]}^R$ in the inserts of figure
\ref{fig:aecuni10} for $\theta = 0.1$ and figure \ref{fig:aeStudyCuNi}
for $\theta = 0.01$ correlate with the peaks in the probability of
generating dimers, $\omega_{D[0,0] \rightarrow [1,0]}^R$, as shown in
figure \ref{fig:omegacuni10} for $\theta = 0.1$ and figure
\ref{fig:someOmegasCuNi} for $\theta = 0.01$, respectively. Thus, as
the temperature is increased, a larger fraction of the monomers start
diffusing and, as a result, there are less obstacles and a lower
probability to form dimers. Consequently, the dominance by dimer
diffusion gives away to the dominance by monomer diffusion.

According to figure \ref{fig:omegacuni10}, the biggest difference with
respect to Ni/Cu(111) at low temperature is the strong dominance by
adsorption (no energy barrier), monomer attachment to the islands
($D[0,0] \rightarrow [1,0]$ and $D[0,0] \rightarrow [2,0]$, with
energy barriers of 35 and 31 meV, respectively) and stabilization of
recently-attached monomers ($D[1,0] \rightarrow [2,0]$, 26 meV;
$D[1,0] \rightarrow [2,2]$, 10 meV; and $D[1,0] \rightarrow [3,0]$,
10 meV). Note that, in the Cu/Ni(111) system, the role of concerted
dimer diffusion ($I2$) is less significant, achieving an event
probability of between 0.1 and 2.8\% at high temperatures. Similarly,
at the highest considered temperatures, edge diffusion
($D[2,0] \rightarrow [2,0]$, with
$E_{[2,0] \rightarrow [2,0]}^{k} =268$ meV) becomes appreciable.

Further plots for the event probabilities as a function of temperature
are shown for representative coverages in figures
\ref{fig:someOmegasCuNi} and \ref{fig:someOmegasNiCu}
of the appendix, for for Cu/Ni(111) and for Ni/Cu(111), respectively.
The trend at any coverage is much equivalent to the picture just
presented. In general, the Cu/Ni(111) system is dominated by monomer
diffusion ($D[0,0] \rightarrow [0,0]$) at high temperatures,
non-concerted dimer diffusion ($D[1,0] \rightarrow [1,0]$) at
intermediate temperatures, and adsorption at low temperatures. At this
end (low temperature), monomer attachment to the islands
($D[0,0] \rightarrow [1,0]$, $D[0,0] \rightarrow [2,0]$ and
$D[0,0] \rightarrow [3,0]$) and stabilisation of recently-attached
monomers ($D[1,0] \rightarrow [2,0]$, $D[1,0] \rightarrow [2,2]$ and
$D[1,0] \rightarrow [3,0]$) are also relevant, becoming more important
the lower the temperature and the higher the coverage, each one on a
different scale. The same trend is valid for the Ni/Cu(111) system,
although the importance of adsorption, monomer attachment and
recently-attached-monomer stabilisation at low temperatures is less
significant. In addition, concerted dimer diffusion has an appreciable
role in this system at all coverages and over the whole range of
temperature. For completeness, the event probabilities for the most
relevant events are also shown as three-dimensional plots against
coverage and inverse temperature in figures \ref{fig:omegas3dcuni} and
\ref{fig:omegas3dnicu} of the appendix, for for Cu/Ni(111) and for
Ni/Cu(111), respectively.


\section{Conclusions}
\label{sec:conclusions}
We perform kinetic Monte Carlo simulations of two-dimensional
submonolayer growth at constant deposition flux, where the rate of
interest---the tracer diffusivity---is shown to be proportional to the
total diffusion rate,
$R_{d} = \sum_{ \alpha \in \{ d \} } M_\alpha k_\alpha$, and closely
related to the total rate,
$R = \sum_{ \alpha \in \{ e \} } M_\alpha k_\alpha$. This means that
the growth process depends on both the rates of the distinct events,
$k_\alpha$, and their multiplicities, $M_\alpha$, i.e.\ the numbers of
locations where each event $\alpha$ can be performed in a given
snapshot of the surface.  Based on this, we focus on the study of two
specific metallic heteroepitaxial systems, namely, Cu on Ni(111) and
Ni on Cu(111), as a function of coverage and temperature while
including a large variety of single-atom, multi-atom and
complete-island diffusion events. The interaction between the atoms is
described with a many body semi empirical embedded atom model and the
drag method is used to calculate the energy barriers. The two systems
are compared in terms of their temperature-dependent morphology,
island density and diffusivity, through the total rates, $R_d$ and
$R$, including their apparent activation energies, $E_{app}^{R}$ and
$E_{app}^{R_d}$.

The use of the multiplicities allows describing the probability of
every event with respect to all others. As a result, we conclude that,
at low temperature, the diffusivity is dominated by dimer diffusion,
which is split between non-concerted dimer diffusion and concerted
dimer diffusion. At medium temperature, it is controlled by monomer
diffusion and, at high temperature, it is due to a mixture of monomer
diffusion, perimeter diffusion and concerted dimer/trimer
diffusion. Thus, this work shows the importance of some concerted
diffusion events in 2D submonolayer epitaxial growth. Although
concerted diffusion has a substantial role in one of the two analysed
systems, it is to be expected that concerted motion may be even more
important in other systems, including the relatively unexplored area
of on-surface synthesis.

Most importantly, the use of the multiplicities enables formulating
the apparent activation energy as a weighted average, where the
weights are identified as the probabilities of the different events
and the actual energy contribution for every event contains both the
traditional energy barrier and an additional unbounded configurational
term, directly related to the temperature dependence of its
multiplicity. Since the leading event in the weighted average may
easily change with the growth conditions and the configurational terms
may vary widely, we show that a constant value of the apparent
activation energy can be obtained even if control shifts from one
elementary reaction to another. This means that the traditional
assignment of a constant apparent activation energy to an underlying
rate determining step is not the only possibility and, thus, it is not
necessarily valid during epitaxial growth.

The study demonstrates that the multiplicity analysis can be applied
for systems with hundreds of distinct events, showing that eventually
a few of them dominate the growth process. In the future, the addition
of self-learning KMC (SLKMC) techniques should enable finding and
executing new diffusion events, for any type of single-atom and
multi-atom event. The present work opens the door to include the
multiplicity analysis into the existing SLKMC methods.

\section*{Acknowledgements}
We acknowledge support by the project ``Connecting Mesoscale Dynamics
of Metallic Films on Semiconductors to Nanoscale Phenomena,'' by the
US National Science Foundation, by the Basque Departamento de
Educacion, UPV/EHU (Grant No. IT-756-13), and the 2015/01 contract by
the DIPC.  The KMC calculations were performed on the ATLAS
supercomputer in the DIPC.  We would like to acknowledge STOKES
advanced computing center at the University of Central Florida for
resources to calculate barriers of processes.  We are thankful to
Dr. N. Ferrando from Universitat Polit\`{e}cnica de Val\`{e}ncia, for
the main development of the PSD tool as well as an initial
implementation of the ``Morphokinetics'' KMC code.
\section*{References}
\bibliography{shorttitles,biblio} 
\bibliographystyle{elsarticle-num}

\cleardoublepage
\newcommand*{\mycommand}[1]{\texttt{\emph{#1}}}
\setcounter{page}{1}
\setcounter{section}{0}
\setcounter{figure}{0}
\renewcommand{\thepage}{A\arabic{page}} 
\renewcommand{\thesection}{A\arabic{section}}  
\renewcommand{\thetable}{A\arabic{table}}  
\renewcommand{\thefigure}{A\arabic{figure}}

\onecolumn
\cleardoublepage
\section{Diffusion events appendix}

\subsection{Single-atom available diffusion events}
\begin{figure}[htb!]
  \begin{center}
    \begin{sideways}
      \def\svgwidth{195mm}
      {\small
\executeiffilenewer{lattice_detailed_atom_types_stabilised.svg}{lattice_detailed_atom_types_stabilised.pdf}%
{inkscape             -z             -D             --file=lattice_detailed_atom_types_stabilised.svg             %
--export-pdf=lattice_detailed_atom_types_stabilised.pdf             --export-latex}%
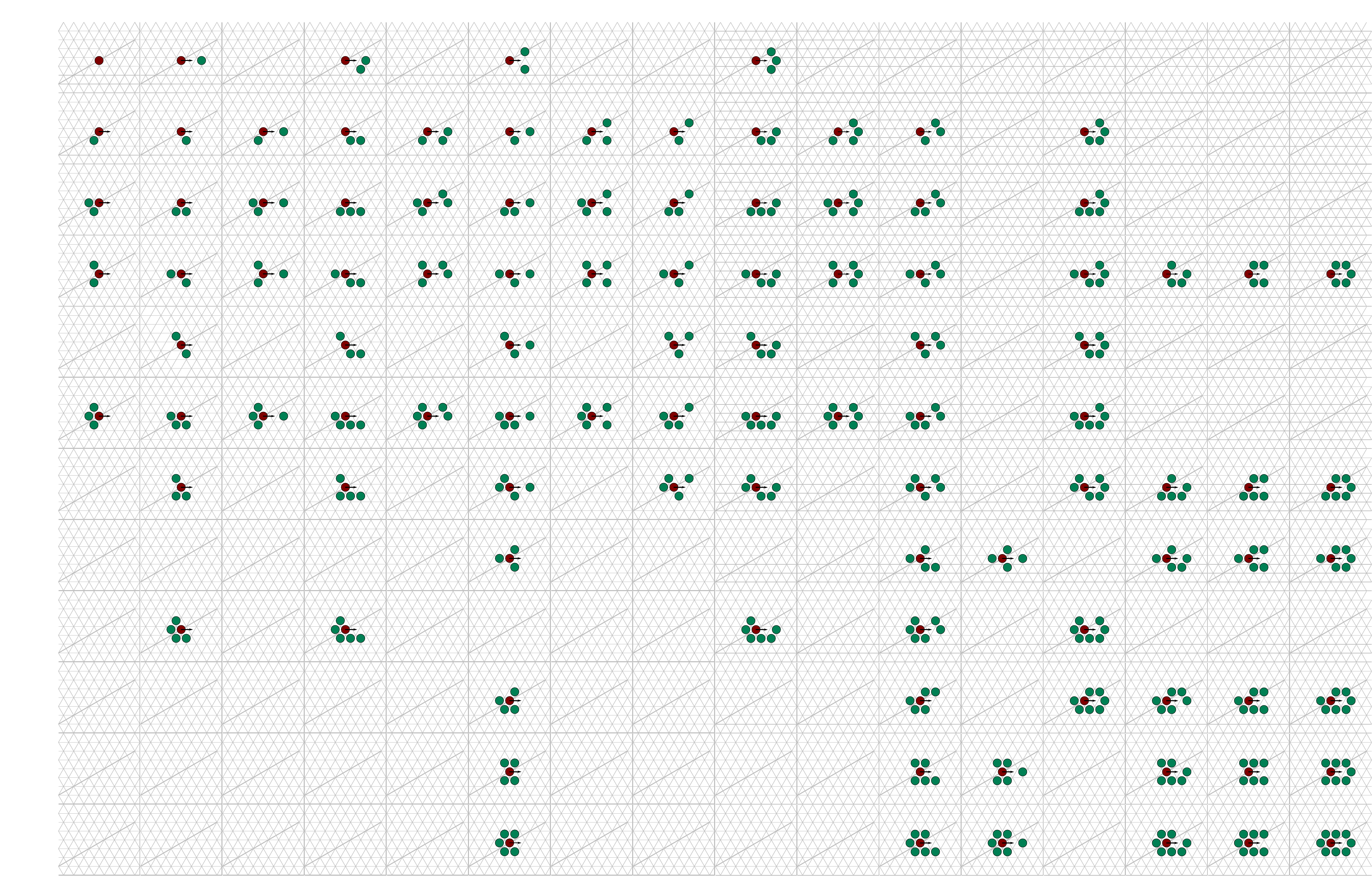%

      }
    \end{sideways}
  \end{center}
  \caption{Complete table for all single atom events considered in this study. 
    The red adatom is moved one position to the
    right. The green adatoms are nearest neighbours of the 
    origin/destination sites. White adatoms have been used to stabilise the
    structures in EAM+drag. * denotes detachment, where 
    the destination site has no neighbours in common with the origin site.
    The table contains 107 single atom event types. Taking into
    account concerted island diffusion (7 types, figure
    \ref{fig:islandGeom}), multi-atom diffusion (4 types, figure
    \ref{fig:multiAtomGeom}) and the adsorption event (1 unique event
    type), the KMC simulations use a total of 119 different event types.}
  \label{fig:all_energies}
\end{figure}

\cleardoublepage
\subsection{Activation energies for Cu/Ni and Ni/Cu}
\begin{table*}[!htb]
  \caption{Microscopic diffusion activation energy barriers (in eV) for Cu/Ni system.}
  \label{tb:cuni}
  {\small
    \bgroup
    \setlength\tabcolsep{1.0mm}
    \begin{tabular}{r|llllllllllllllll}
      \multicolumn{1}{r}{Type} & \multicolumn{3}{l}{ \scriptsize{[class, subclass]}} &   \multicolumn{12}{c}{Single atom diffusion events} \\ \hline
                               & 0 \scriptsize{[0,0]} & 1 \scriptsize{[1,0]} & 2 \scriptsize{[1,0]$^*$} & 3 \scriptsize{[2,0]} & 4 \scriptsize{[2,0]$^*$} & 5 \scriptsize{[2,1]} & 6 \scriptsize{[2,1]$^*$} & 7 \scriptsize{[2,2]} & 8 \scriptsize{[3,0]} & 9 \scriptsize{[3,0]$^*$} & 10 \scriptsize{[3,1]} & 11 \scriptsize{[3,2]} & 12 \scriptsize{[4,0]} & 13 \scriptsize{[4,1]} & 14 \scriptsize{[4,2]} & 15 \scriptsize{[5,0]} \\ \hline

0 \scriptsize{[0,0]}   & 0.052                & 0.044                & $\times$                 & 0.029                & $\times$                 & 0.005                & $\times$                 & $\times$             & 0.0024               & $\times$                 & $\times$              & $\times$              & $\times$              & $\times$              & $\times$              & $\times$              \\
1  \scriptsize{[1,0]}  & 0.428                & 0.038                & 0.317                    & 0.026                & 0.258                    & 0.033                & 0.183                    & 0.0027               & 0.01                 & 0.184                    & 0.0012                & $\times$              & 0.00086               & $\times$              & $\times$              & $\times$              \\
2 \scriptsize{[2,0]}   & 0.736                & 0.360                & 0.625                    & 0.268                & 0.433                    & 0.261                & 0.383                    & 0.167                & 0.22                 & 0.396                    & 0.164                 & $\times$              & 0.144                 & $\times$              & $\times$              & $\times$              \\
3  \scriptsize{[2,1]}  & 0.750                & 0.397                & 0.565                    & 0.308                & 0.206                    & 0.293                & 0.167                    & 0.197                & 0.258                & 0.179                    & 0.185                 & 0.651                 & 0.176                 & 0.503                 & 0.439                 & 0.430                 \\
4 \scriptsize{[2,2]}   & $\times$             & 0.403                & $\times$                 & 0.198                & $\times$                 & 0.189                & $\times$                 & 0.373                & 0.413                & $\times$                 & 0.328                 & $\times$              & 0.292                 & $\times$              & $\times$              & $\times$              \\
5  \scriptsize{[3,0]}  & 1.010                & 0.663                & 0.828                    & 0.546                & 0.483                    & 0.539                & 0.413                    & 0.390                & 0.473                & 0.400                    & 0.369                 & $\times$              & 0.357                 & $\times$              & $\times$              & $\times$              \\
6  \scriptsize{[3,1]}  & $\times$             & 0.697                & $\times$                 & 0.502                & $\times$                 & 0.479                & $\times$                 & 0.360                & 0.386                & $\times$                 & 0.188                 & 0.905                 & 0.184                 & 0.804                 & 0.683                 & 0.615                 \\
7  \scriptsize{[3,2]}  & $\times$             & $\times$             & $\times$                 & $\times$             & $\times$                 & 0.899                & $\times$                 & $\times$             & $\times$             & $\times$                 & 0.748                 & 1.01                  & $\times$              & 0.841                 & 0.687                 & 0.726                 \\
8   \scriptsize{[4,0]} & $\times$             & 0.947                & $\times$                 & 0.748                & $\times$                 & 0.957                & $\times$                 & 0.851                & 0.627                & $\times$                 & 0.448                 & $\times$              & 0.423                 & $\times$              & $\times$              & $\times$              \\
9  \scriptsize{[4,1]}  & $\times$             & $\times$             & $\times$                 & $\times$             & $\times$                 & 1.010                & $\times$                 & $\times$             & $\times$             & $\times$                 & 0.895                 & 1.107                 & 0.763                 & 0.855                 & 0.813                 & 0.763                 \\
10 \scriptsize{[4,2]}  & $\times$             & $\times$             & $\times$                 & $\times$             & $\times$                 & 1.020                & $\times$                 & $\times$             & $\times$             & $\times$                 & 0.815                 & 0.964                 & $\times$              & 0.807                 & 0.733                 & 0.729                 \\
11  \scriptsize{[5,0]} & $\times$             & $\times$             & $\times$                 & $\times$             & $\times$                 & 1.144                & $\times$                 & $\times$             & $\times$             & 1.010                    & 1.152                 & 1.008                 & $\times$              & 0.904                 & 0.908                 & 0.908                 \\
  \vspace{3mm}
    \end{tabular}
    \egroup
  }
  \centering
  \vspace{3mm}
  \begin{tabular}{cc}
    \begin{tabular}{rrll}
      Concerted island diffusion &      &                        \\ \hline
      Atoms in the island        & Type & Energy (eV) & Name     \\ \hline
      2                          & I2   & 0.062       & Dimer    \\
      3                          & I3   & 0.161       & Trimer   \\
      4                          & I4   & 0.182       & Tetramer \\
      5                          & I5   & 0.222       & Pentamer \\
      6                          & I6   & 0.201       & Hexamer  \\
      7                          & I7   & 0.403       & Heptamer \\
      8                          & I8   & 0.372       & Octamer          
    \end{tabular}
                                 & 
                            
                            \begin{tabular}{rrl}
                              \multicolumn{2}{c}{Concerted two-atom diffusion}              \\ \hline
                              Multi-atom type              & Energy (eV) \\ \hline
                              C1                            & 0.481       \\
                              C2                            & 0.437       \\
                              C3                            & 0.397       \\
                              C4                            & 0.228
                            \end{tabular}
  \end{tabular}
\end{table*}

\begin{table*}[htb!]
  \caption{Microscopic diffusion activation energy barriers (in eV) for Ni/Cu system.}
  \label{tb:nicu}
  {\small
    \bgroup
    \setlength\tabcolsep{1.0mm}
    \begin{tabular}{r|llllllllllllllll}
      \multicolumn{1}{r}{Type} & \multicolumn{3}{l}{ \scriptsize{[class, subclass]}} &   \multicolumn{12}{c}{Single atom diffusion events} \\ \hline
                               & 0 \scriptsize{[0,0]} & 1 \scriptsize{[1,0]} & 2 \scriptsize{[1,0]$^*$} & 3 \scriptsize{[2,0]} & 4 \scriptsize{[2,0]$^*$} & 5 \scriptsize{[2,1]} & 6 \scriptsize{[2,1]$^*$} & 7 \scriptsize{[2,2]} & 8 \scriptsize{[3,0]} & 9 \scriptsize{[3,0]$^*$} & 10 \scriptsize{[3,1]} & 11 \scriptsize{[3,2]} & 12 \scriptsize{[4,0]} & 13 \scriptsize{[4,1]} & 14 \scriptsize{[4,2]} & 15 \scriptsize{[5,0]} \\ \hline
      0 \scriptsize{[0,0]}       & 0.031    & 0.028    & $\times$ & 0.015    & $\times$ & 0.009 & $\times$ & $\times$ & 0.008    & $\times$ & $\times$ & $\times$ & $\times$ & $\times$ & $\times$ & $\times$ \\
      1  \scriptsize{[1,0]}      & 0.568    & 0.016    & 0.505    & 0.014    & 0.159    & 0.006 & 0.172    & 0.000    & 0.001    & 0.180    & 0.000    & $\times$ & 0.023    & $\times$ & $\times$ & $\times$ \\
      2 \scriptsize{[2,0]}       & 0.938    & 0.439    & 0.746    & 0.364    & 0.743    & 0.389 & 0.541    & 0.305    & 0.500    & 0.562    & 0.319    & $\times$ & 0.263    & $\times$ & $\times$ & $\times$ \\
      3  \scriptsize{[2,1]}      & 0.800    & 0.489    & 0.550    & 0.450    & 0.467    & 0.448 & 0.356    & 0.366    & 0.404    & 0.372    & 0.382    & 0.659    & 0.353    & 0.531    & 0.423    & 0.643    \\
      4 \scriptsize{[2,2]}       & $\times$ & 0.678    & $\times$ & 0.340    & $\times$ & 0.334 & $\times$ & 0.596    & 0.283    & $\times$ & 0.115    & $\times$ & 0.170    & $\times$ & $\times$ & $\times$ \\
      5  \scriptsize{[3,0]}      & 1.290    & 0.804    & $\infty$ & 0.704    & 0.500    & 0.742 & 0.662    & 0.629    & 0.644    & 0.658    & 0.645    & $\times$ & 0.571    & $\times$ & $\times$ & $\times$ \\
      6  \scriptsize{[3,1]}      & $\times$ & 0.875    & $\times$ & 0.690    & $\times$ & 0.693 & $\times$ & 0.482    & 0.57     & $\times$ & 0.518    & 0.902    & 0.429    & 0.892    & 0.931    & 0.945    \\
      7  \scriptsize{[3,2]}      & $\times$ & $\times$ & $\times$ & $\times$ & $\times$ & 1.162 & $\times$ & $\times$ & $\times$ & $\times$ & 1.144    & 1.191    & $\times$ & 1.129    & 1.259    & 0.992    \\
      8   \scriptsize{[4,0]}     & $\times$ & 1.145    & $\times$ & 0.959    & $\times$ & 1.220 & $\times$ & 1.157    & 0.858    & $\times$ & 0.839    & $\times$ & 0.726    & $\times$ & $\times$ & $\times$ \\
      9  \scriptsize{[4,1]}      & $\times$ & $\times$ & $\times$ & $\times$ & $\times$ & 1.330 & $\times$ & $\times$ & $\times$ & $\times$ & 1.384    & 1.209    & 1.225    & 1.400    & 1.226    & 1.000    \\
      10 \scriptsize{[4,2]}      & $\times$ & $\times$ & $\times$ & $\times$ & $\times$ & 1.310 & $\times$ & $\times$ & $\times$ & $\times$ & 1.120    & 1.291    & $\times$ & 1.090    & 1.175    & 1.100    \\
      11  \scriptsize{[5,0]}     & $\times$ & $\times$ & $\times$ & $\times$ & $\times$ & 1.507 & $\times$ & $\times$ & $\times$ & 1.382    & 1.482    & 1.326    & $\times$ & 1.326    & 1.361    & 1.174
    \vspace{3mm}
    \end{tabular}
    \egroup
  }
  \vspace{3mm}
  \centering
  \begin{tabular}{cc}
    \begin{tabular}{rrll}
      Concerted island diffusion &      &             &          \\ \hline
      Atoms in the island        & Type & Energy (eV) & Name     \\ \hline
      2                          & I2   & 0.021       & Dimer    \\
      3                          & I3   & 0.148       & Trimer   \\
      4                          & I4   & 0.157       & Tetramer \\
      5                          & I5   & 0.220       & Pentamer \\
      6                          & I6   & 0.199       & Hexamer  \\
      7                          & I7   & 0.369       & Heptamer \\
      8                          & I8   & 0.380       & Octamer          
    \end{tabular}
                                 & 
                            
                            \begin{tabular}{rrl}
                              \multicolumn{2}{c}{Concerted two-atom diffusion}      \\ \hline
                              Multi-atom                   & Energy (eV) \\ \hline
                               C1                          & 0.654       \\
                               C2                          & 0.633       \\
                               C3                          & 0.294       \\
                               C4                          & 0.218
                            \end{tabular}
  \end{tabular}
\end{table*}
\cleardoublepage
\setcounter{page}{1}
\setcounter{section}{0}
\setcounter{figure}{0}
\renewcommand{\thepage}{B\arabic{page}} 
\renewcommand{\thesection}{B\arabic{section}}  
\renewcommand{\thetable}{B\arabic{table}}  
\renewcommand{\thefigure}{B\arabic{figure}}
\restoregeometry
\onecolumn
\section{Results appendix}

\subsection{Additional plots for the total rate and its apparent activation energy}
\label{sec:extraapp}

\begin{figure*}[htb!]
  \begin{subfigure}{0.5\columnwidth}
    \centering
    \caption{Cu/Ni(111)}
    \label{fig:cunitotalrates}
    \def\svgwidth{\textwidth}
\executeiffilenewer{cuNiManyTotalRates.svg}{cuNiManyTotalRates.pdf}%
{inkscape             -z             -D             --file=cuNiManyTotalRates.svg             %
--export-pdf=cuNiManyTotalRates.pdf             --export-latex}%
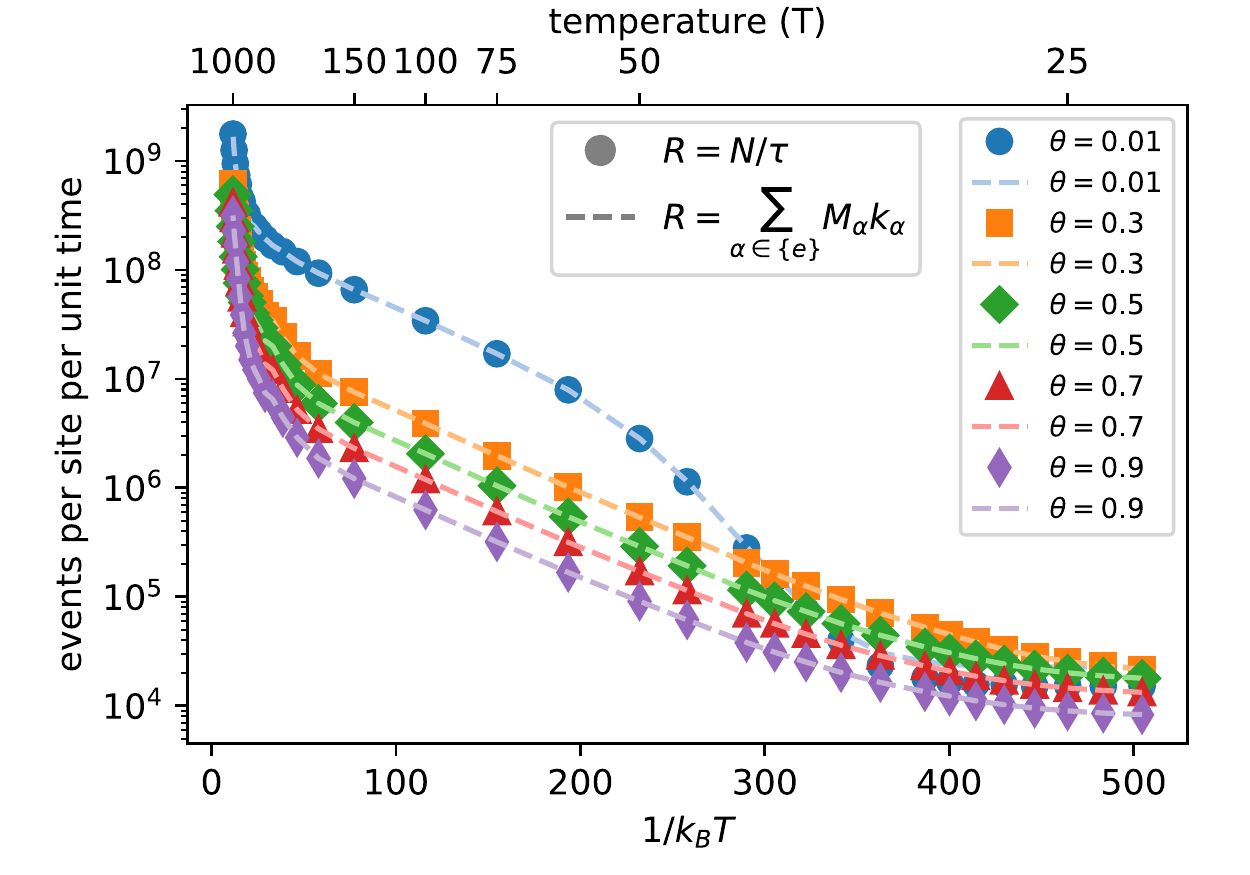%

  \end{subfigure}
  \begin{subfigure}{0.5\columnwidth}
    \centering
    \caption{Ni/Cu(111)}
    \label{fig:nicutotalrates}
    \def\svgwidth{\textwidth}
\executeiffilenewer{niCuManyTotalRates.svg}{niCuManyTotalRates.pdf}%
{inkscape             -z             -D             --file=niCuManyTotalRates.svg             %
--export-pdf=niCuManyTotalRates.pdf             --export-latex}%
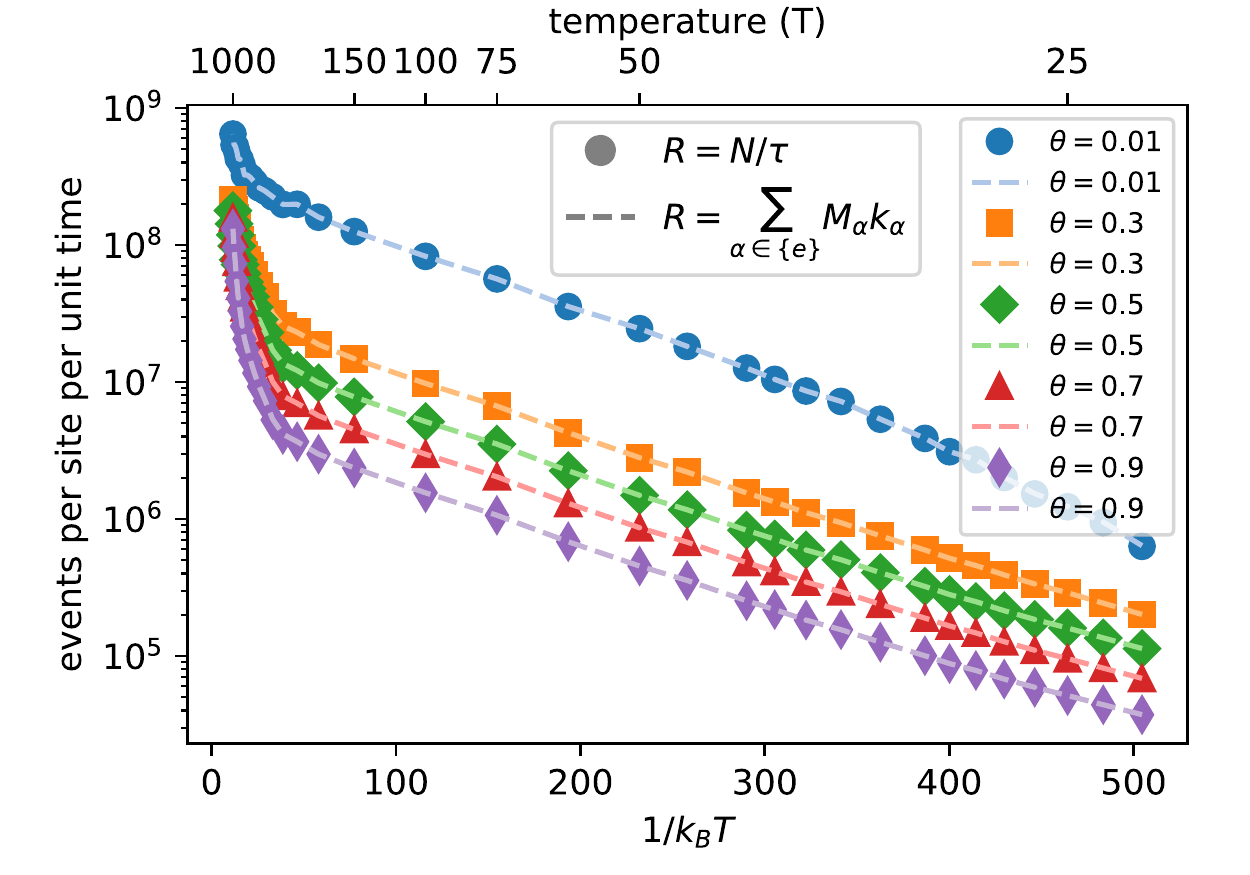%

  \end{subfigure}
  \caption{Average total rate per site, $R = R_{d} + R_{a}$, as a
    function of inverse temperature for various coverages, as
    indicated, for (\subref{fig:cunitotalrates}) Cu/Ni(111), and
    (\subref{fig:nicutotalrates}) Ni/Cu(111).}
  \label{fig:allTotalRates}
\end{figure*}

\begin{figure*}[htb!]
  \begin{subfigure}{0.5\columnwidth}
    \centering
    \caption{Cu/Ni(111)}
    \label{fig:cuNiAllTotalRatesMap}
    \def\svgwidth{\textwidth}
\executeiffilenewer{cuniTotalRateMap.svg}{cuniTotalRateMap.pdf}%
{inkscape             -z             -D             --file=cuniTotalRateMap.svg             %
--export-pdf=cuniTotalRateMap.pdf             --export-latex}%
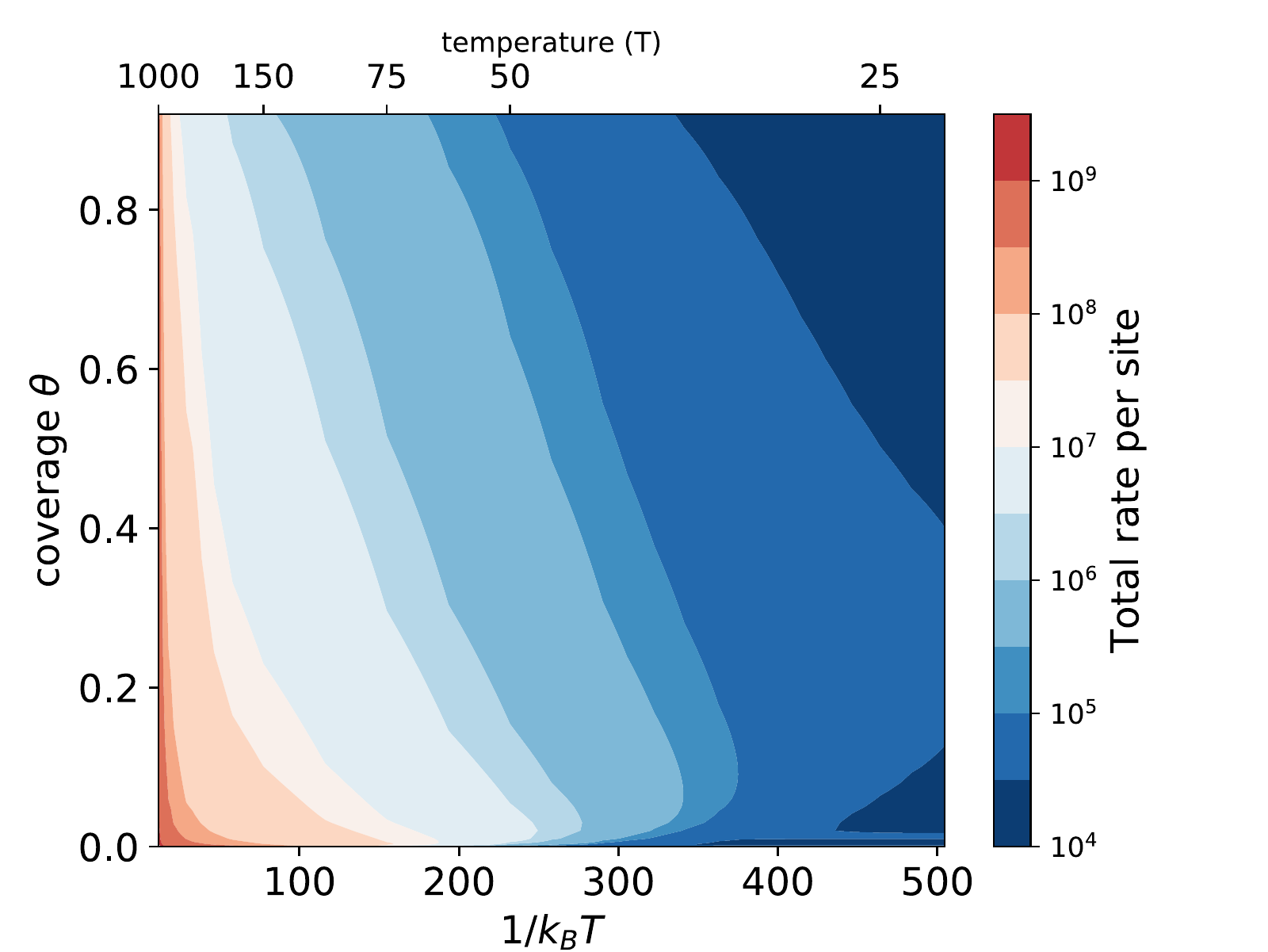%

  \end{subfigure}
  \begin{subfigure}{0.5\columnwidth}
    \centering
    \caption{Ni/Cu(111)}
    \label{fig:niCuAllTotalRatesMap}
    \def\svgwidth{\textwidth}
\executeiffilenewer{nicuTotalRateMap.svg}{nicuTotalRateMap.pdf}%
{inkscape             -z             -D             --file=nicuTotalRateMap.svg             %
--export-pdf=nicuTotalRateMap.pdf             --export-latex}%
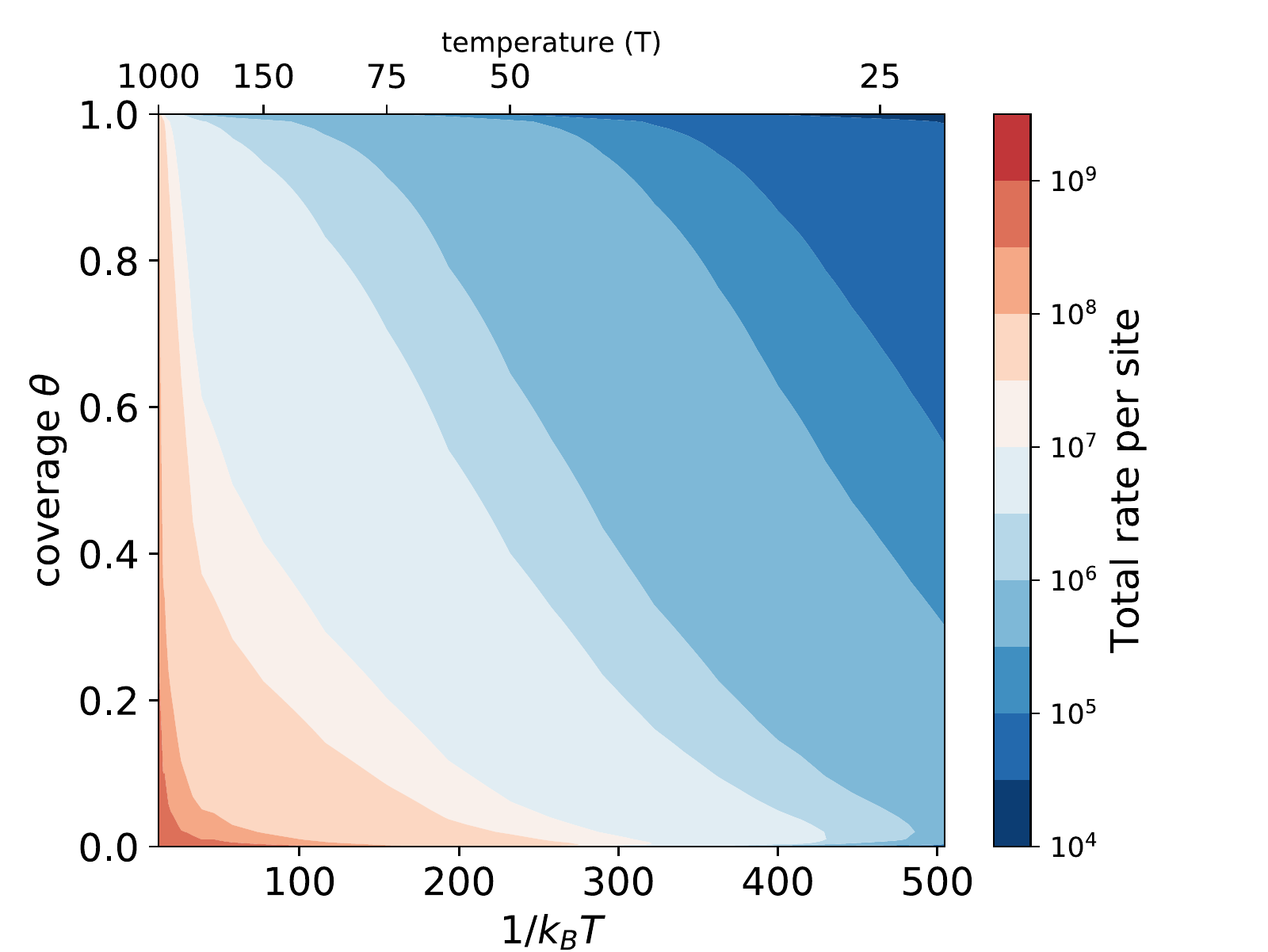%

  \end{subfigure}
  \caption{Top view of a three-dimensional plot of the average total
    rate per site, $R = R_{d} + R_{a}$, as a function of both coverage
    and temperature/inverse temperature for
    (\subref{fig:cuNiAllTotalRatesMap}) Cu/Ni(111), and
    (\subref{fig:niCuAllTotalRatesMap}) Ni/Cu(111).}
  \label{fig:allTotalRatesMap}
\end{figure*}

\begin{figure*}[htb!]
  \begin{subfigure}{0.5\columnwidth}
    \centering
    \caption{Cu/Ni(111)}
    \label{fig:allAeCuNi}
    \def\svgwidth{\textwidth}
\executeiffilenewer{cuniSlopesMap.svg}{cuniSlopesMap.pdf}%
{inkscape             -z             -D             --file=cuniSlopesMap.svg             %
--export-pdf=cuniSlopesMap.pdf             --export-latex}%
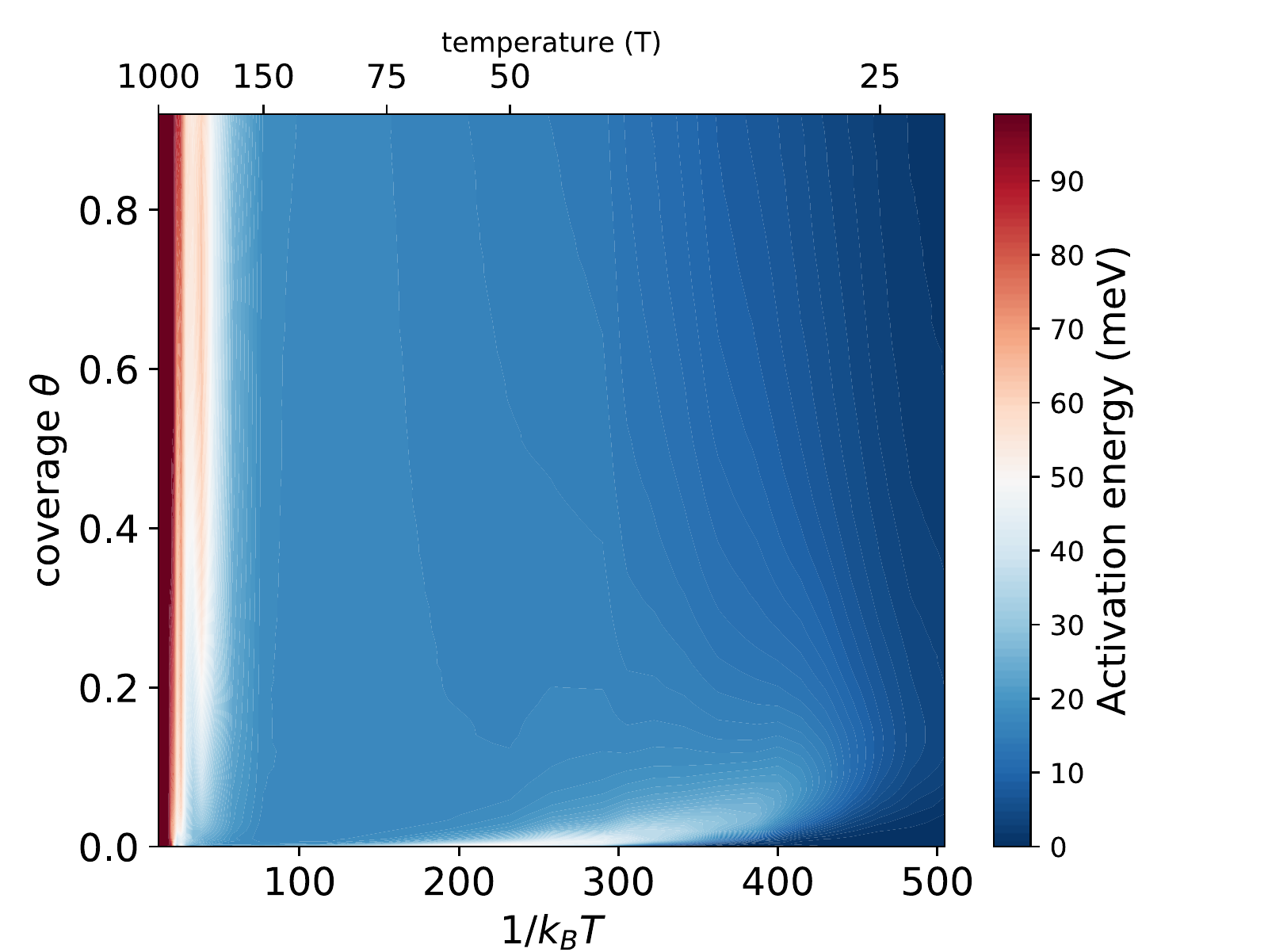%

  \end{subfigure}
  \begin{subfigure}{0.5\columnwidth}
    \centering
    \caption{Ni/Cu(111)}
    \label{fig:allAeNiCu}
    \def\svgwidth{\textwidth}
\executeiffilenewer{nicuSlopesMap.svg}{nicuSlopesMap.pdf}%
{inkscape             -z             -D             --file=nicuSlopesMap.svg             %
--export-pdf=nicuSlopesMap.pdf             --export-latex}%
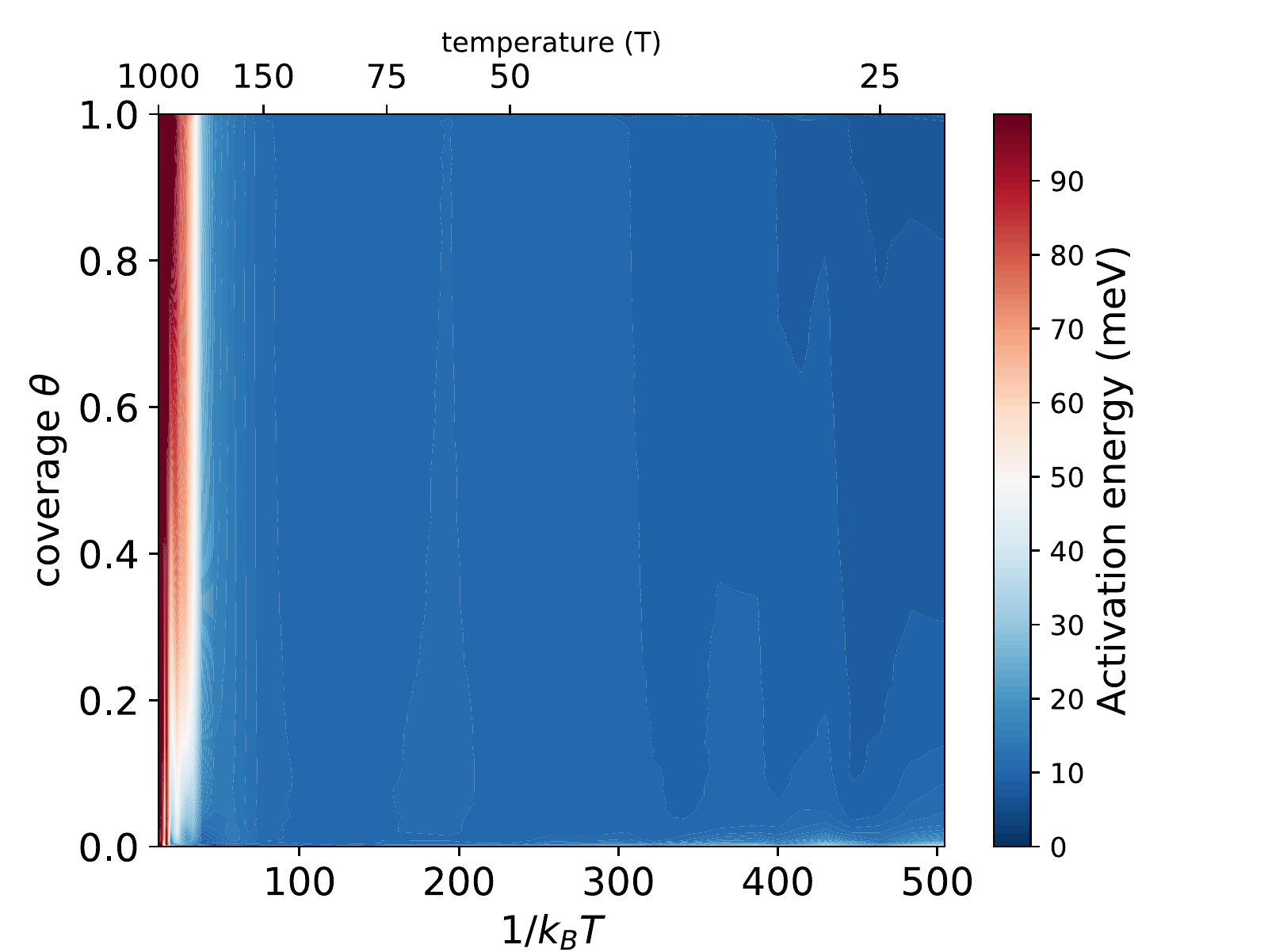%

  \end{subfigure}
  \caption{Top view of a three-dimensional plot of the apparent
    activation energy of the average total rate per site,
    $E_{app}^{R}$, as a function of both coverage and
    temperature/inverse temperature for (\subref{fig:allAeCuNi})
    Cu/Ni(111), and (\subref{fig:allAeNiCu}) Ni/Cu(111).}
  \label{fig:allAe}
\end{figure*}

\begin{figure*}[htb!]
    \centering
    \def\svgwidth{0.9\textwidth}
\executeiffilenewer{dimers.svg}{dimers.pdf}%
{inkscape             -z             -D             --file=dimers.svg             %
--export-pdf=dimers.pdf             --export-latex}%
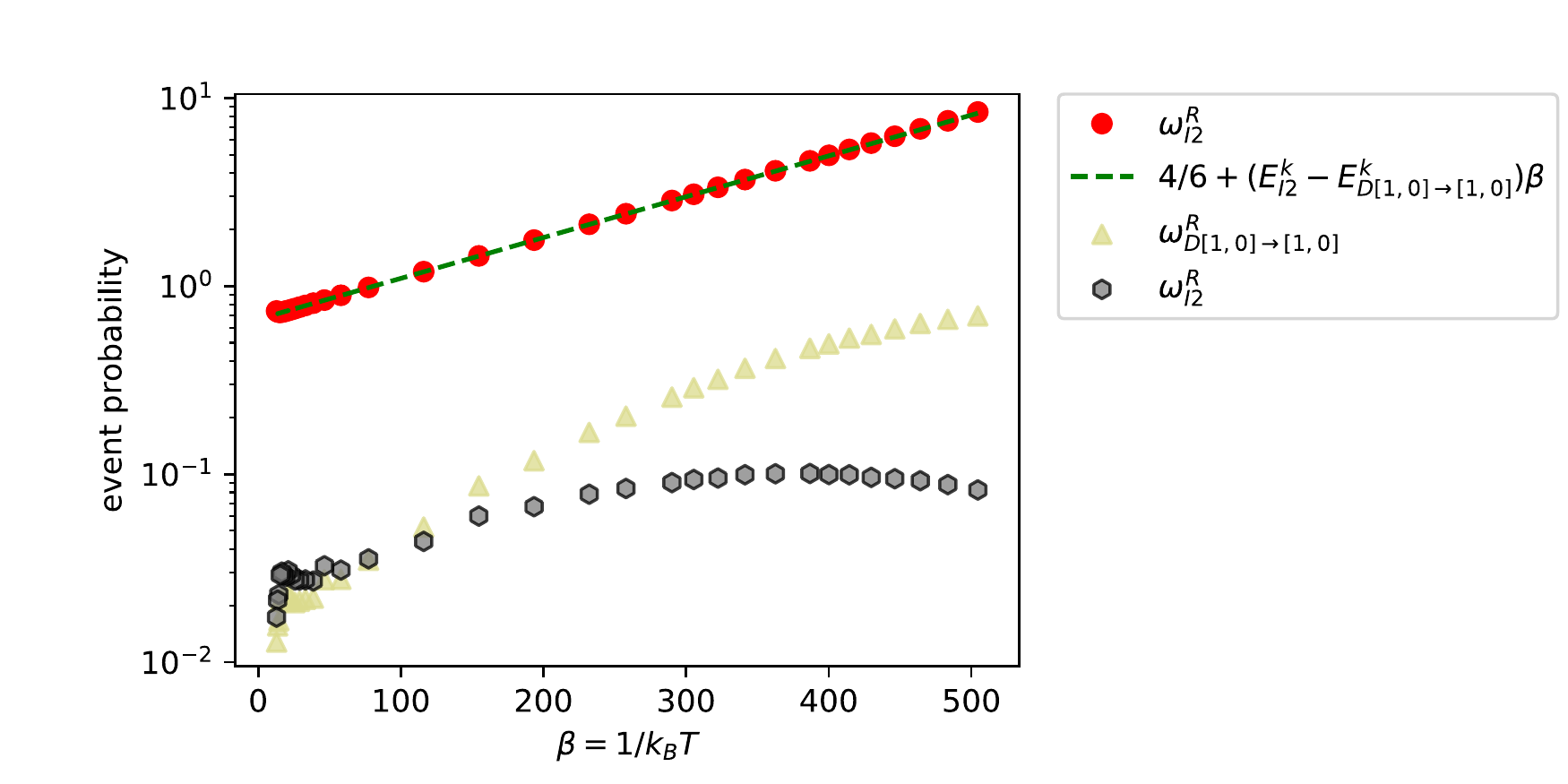%

    \caption{Ratio of the probability to observe non-concerted dimer
      diffusion to that for concerted diffusion,
      $\frac{ \omega_{ D[1,0]\rightarrow[1,0] }^R }{ \omega_{ I2 }^R }
      = \frac{ M_{ D[1,0]\rightarrow[1,0] } k_{ D[1,0]\rightarrow[1,0]
        } }{ M_{ I2 } k_{ I2 } }$ and the expected result
      $\frac{ 4 e^{-E_{ D[1,0]\rightarrow[1,0] }^k \beta} }{ 6 e^{-E_{
            I2 }^k \beta} }$, for Ni/Cu(111) at $\theta = 0.10$. This
      plot confirms that the events $D[1,0]\rightarrow[1,0]$ and $I2$
      correspond to non-concerted dimer diffusion and concerted dimer
      diffusion, respectively. Note that the ratio is larger than 1 at
      low temperatures, indicating that non-concerted dimer diffusion
      is more probable, while the ratio becomes smaller than 1 at high
      temperatures, demonstrating that concerted dimer diffusion
      occurs more frequently.}
  \label{fig:dimers}
\end{figure*}

\begin{figure*}[htb!]
  \centering
  \def\svgwidth{0.9\textwidth}
\executeiffilenewer{cuniSelectionResume.svg}{cuniSelectionResume.pdf}%
{inkscape             -z             -D             --file=cuniSelectionResume.svg             %
--export-pdf=cuniSelectionResume.pdf             --export-latex}%
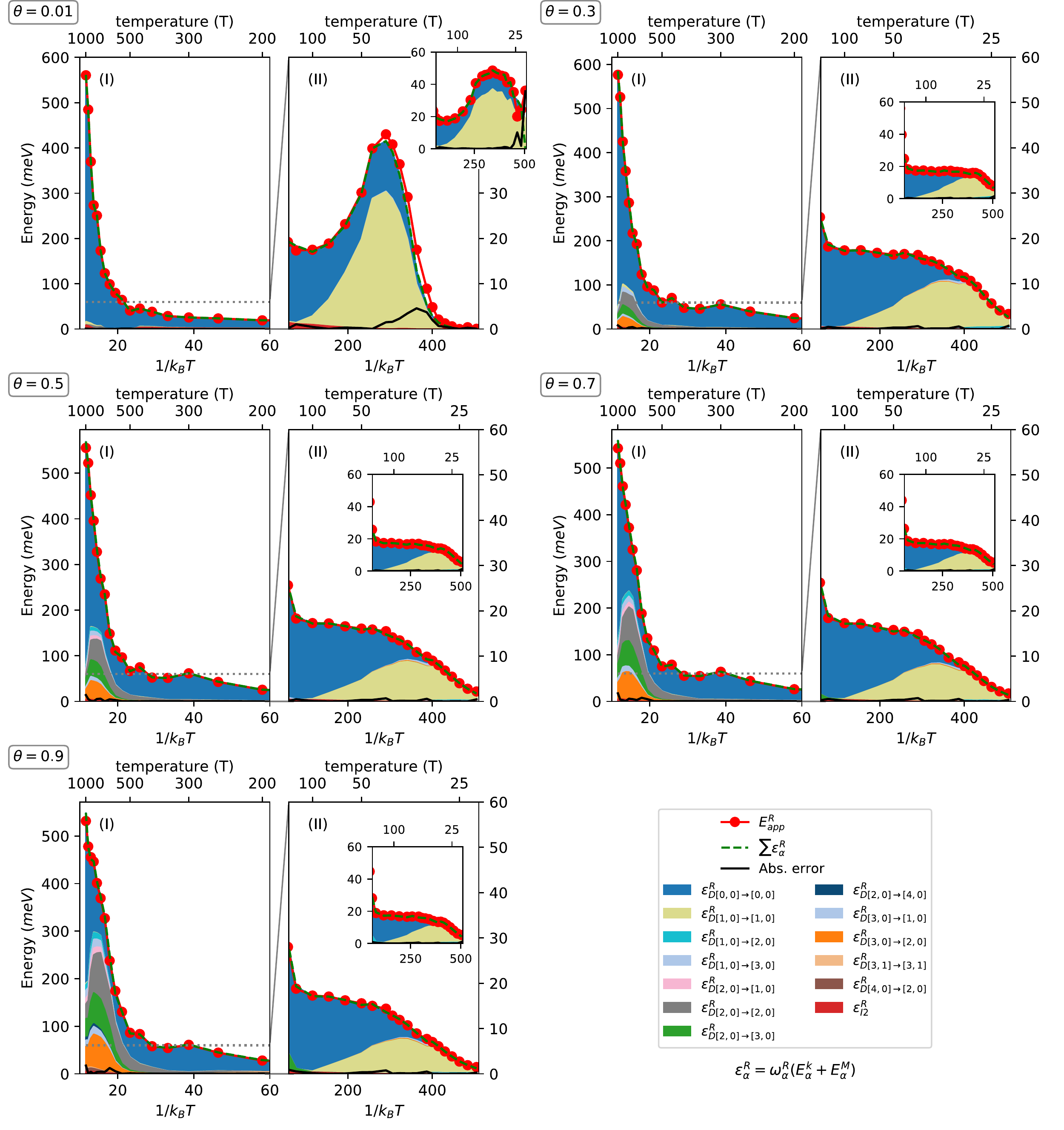%

  \caption{Temperature dependence of $E_{app}^{R}$ for Cu/Ni(111) at
    representative coverage values, as indicated. Two temperature
    regions are shown: (I) $1000 \ge T > 150$ K, and (II)
    $150 \ge T \ge 23$ K, with region II magnified. $E_{app}^{R}$ is
    described well by
    $\sum_{\alpha \in \{ e \}} \epsilon_{\alpha}^{R}$, where
    $\epsilon_{\alpha}^{R} = \omega_{\alpha}^{R} (E_{\alpha}^{k} +
    E_{\alpha}^{M})$. The absolute error
    $| E_{app}^{R} - \sum_{\alpha \in \{ e \}} \epsilon_{\alpha}^{R}
    |$ is also plotted. The number of significant contributions to
    $E_{app}^{R}$ increases with coverage and temperature. The insert
    in region II displays $E_{app}^{R_{d}}$.}
  \label{fig:aeStudyCuNi}
\end{figure*}
\pagebreak
\begin{figure*}[htb!]
  \centering
  \def\svgwidth{0.9\textwidth}
\executeiffilenewer{nicuSelectionResume.svg}{nicuSelectionResume.pdf}%
{inkscape             -z             -D             --file=nicuSelectionResume.svg             %
--export-pdf=nicuSelectionResume.pdf             --export-latex}%
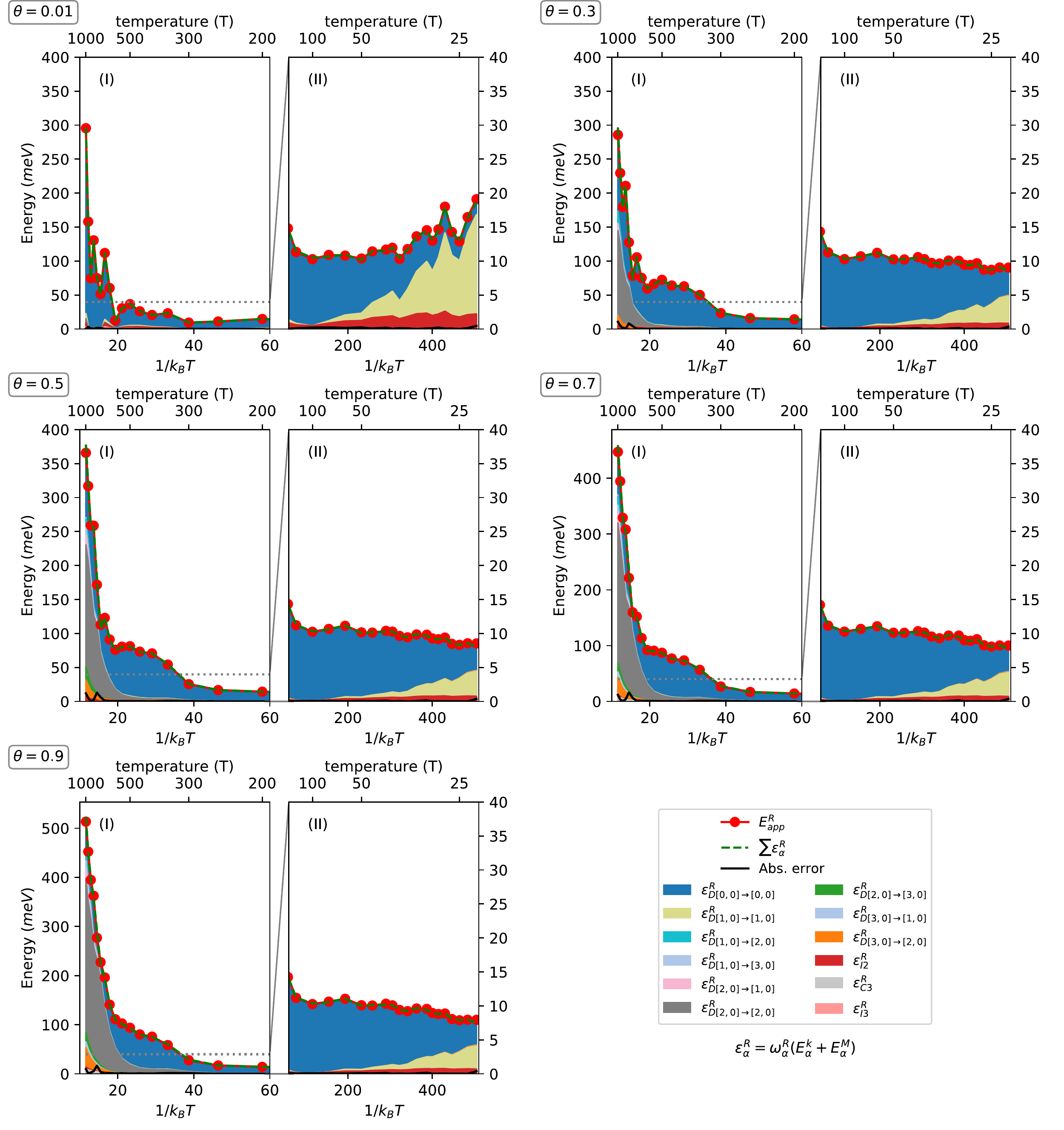%

  \caption{Temperature dependence of $E_{app}^{R}$ for Ni/Cu(111) at
    representative coverage values, as indicated. Two temperature
    regions are shown: (I) $1000 \ge T > 150$ K, and (II)
    $150 \ge T \ge 23$ K, with region II magnified. $E_{app}^{R}$ is
    described well by
    $\sum_{\alpha \in \{ e \}} \epsilon_{\alpha}^{R}$, where
    $\epsilon_{\alpha}^{R} = \omega_{\alpha}^{R} (E_{\alpha}^{k} +
    E_{\alpha}^{M})$. The absolute error
    $| E_{app}^{R} - \sum_{\alpha \in \{ e \}} \epsilon_{\alpha}^{R}
    |$ is also plotted. The number of significant contributions to
    $E_{app}^{R}$ increases with coverage and temperature.}
  \label{fig:aeStudyNiCu}
\end{figure*}

\begin{figure}
  \centering
  \def\svgwidth{0.9\textwidth}
\executeiffilenewer{cuniPlotResume.svg}{cuniPlotResume.pdf}%
{inkscape             -z             -D             --file=cuniPlotResume.svg             %
--export-pdf=cuniPlotResume.pdf             --export-latex}%
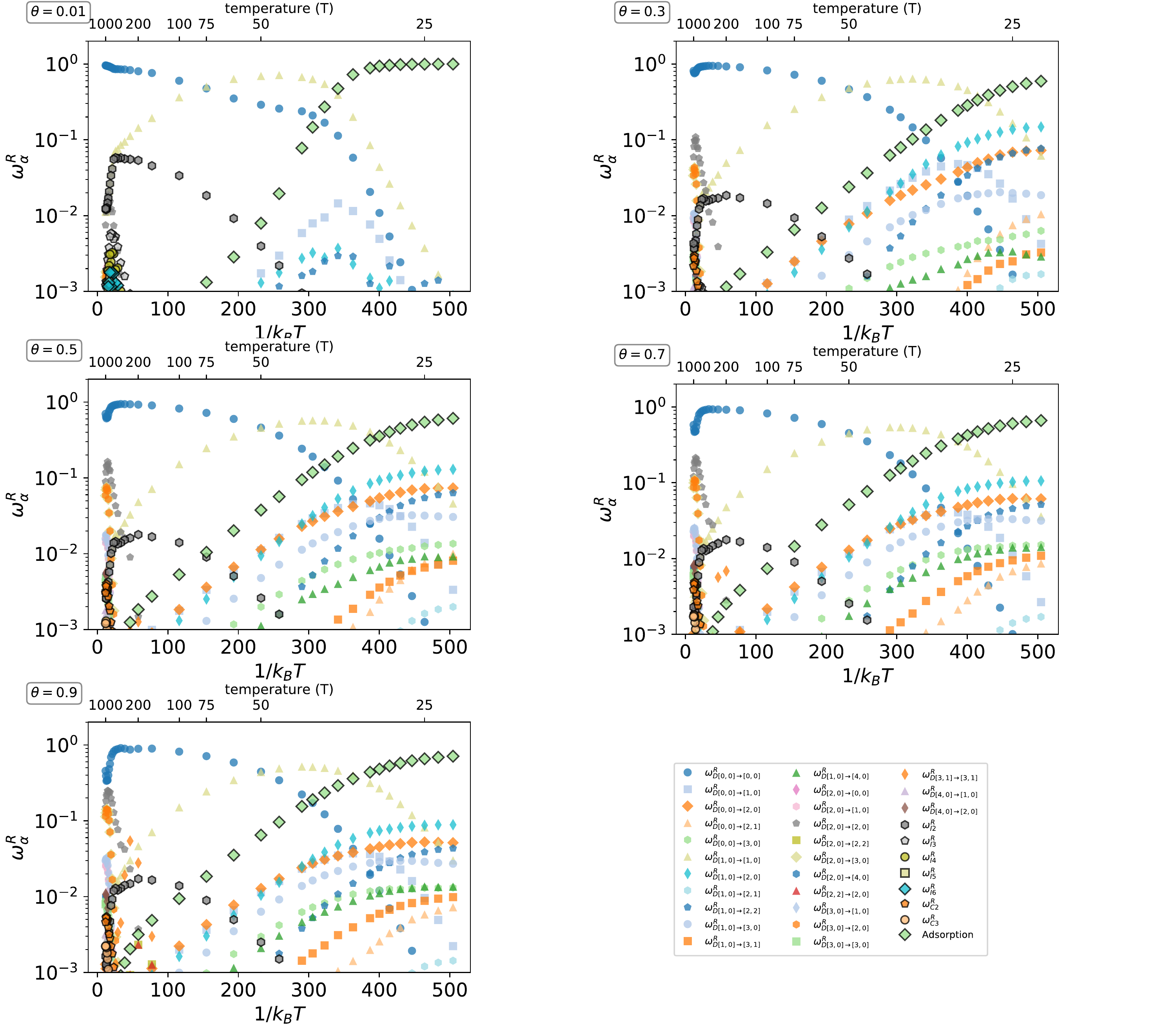%

  \caption{Temperature dependence of the event probabilities ($\omega_{\alpha}^{R}$) for Cu/Ni(111) at representative coverages, as indicated. Only those events whose probability is higher than $10^{-3}$ are shown.}
  \label{fig:someOmegasCuNi}
\end{figure}

\begin{figure}
  \centering
  \def\svgwidth{0.9\textwidth}
\executeiffilenewer{nicuPlotResume.svg}{nicuPlotResume.pdf}%
{inkscape             -z             -D             --file=nicuPlotResume.svg             %
--export-pdf=nicuPlotResume.pdf             --export-latex}%
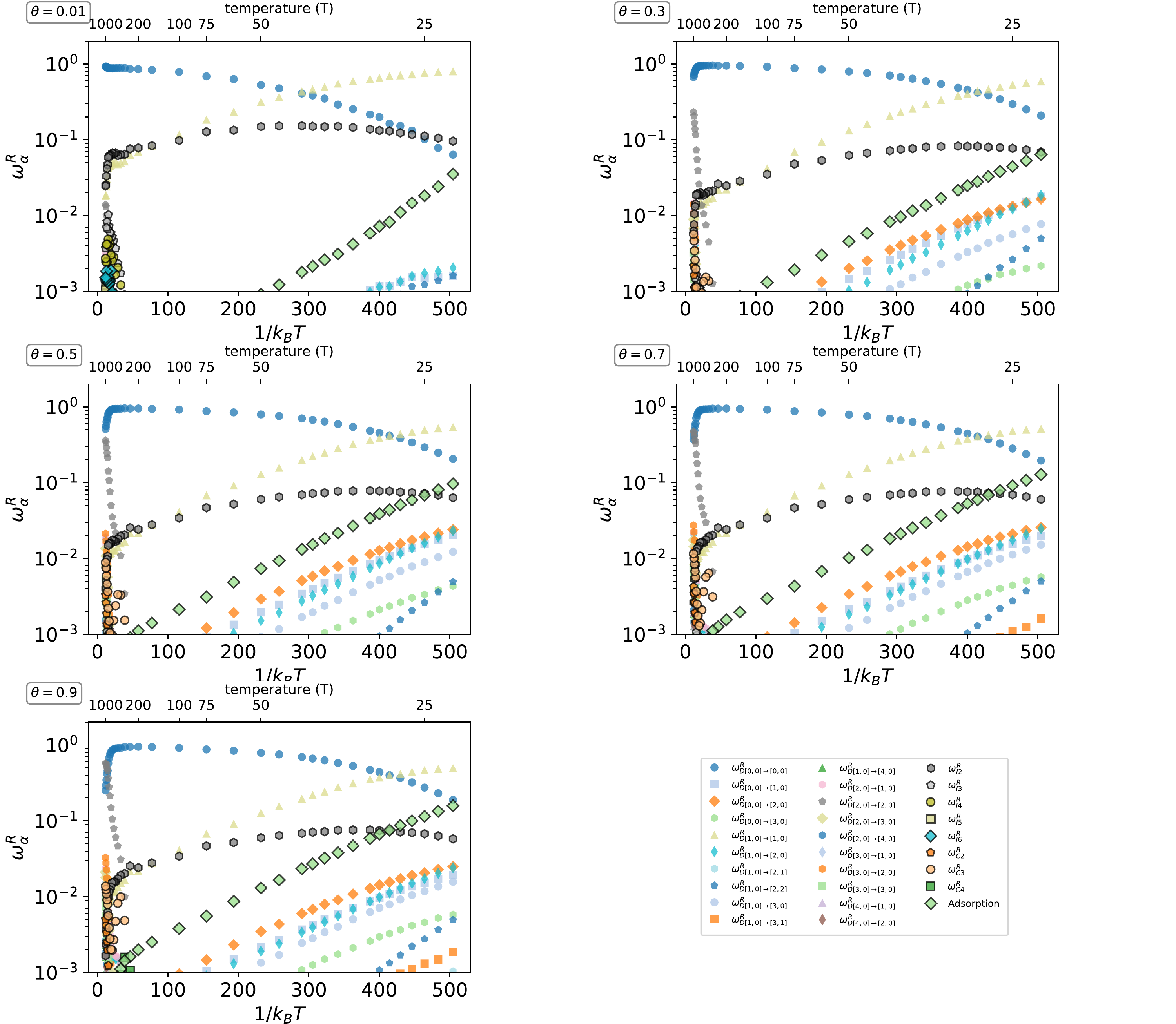%

  \caption{Temperature dependence of the event probabilities
    ($\omega_{\alpha}^{R}$) for Ni/Cu(111) at representative coverages, 
    as indicated. Only those events whose probability is
    higher than $10^{-3}$ are shown.}
  \label{fig:someOmegasNiCu}
\end{figure}

\begin{figure}[htb!]
  \centering
  \includegraphics[width=0.85\columnwidth]{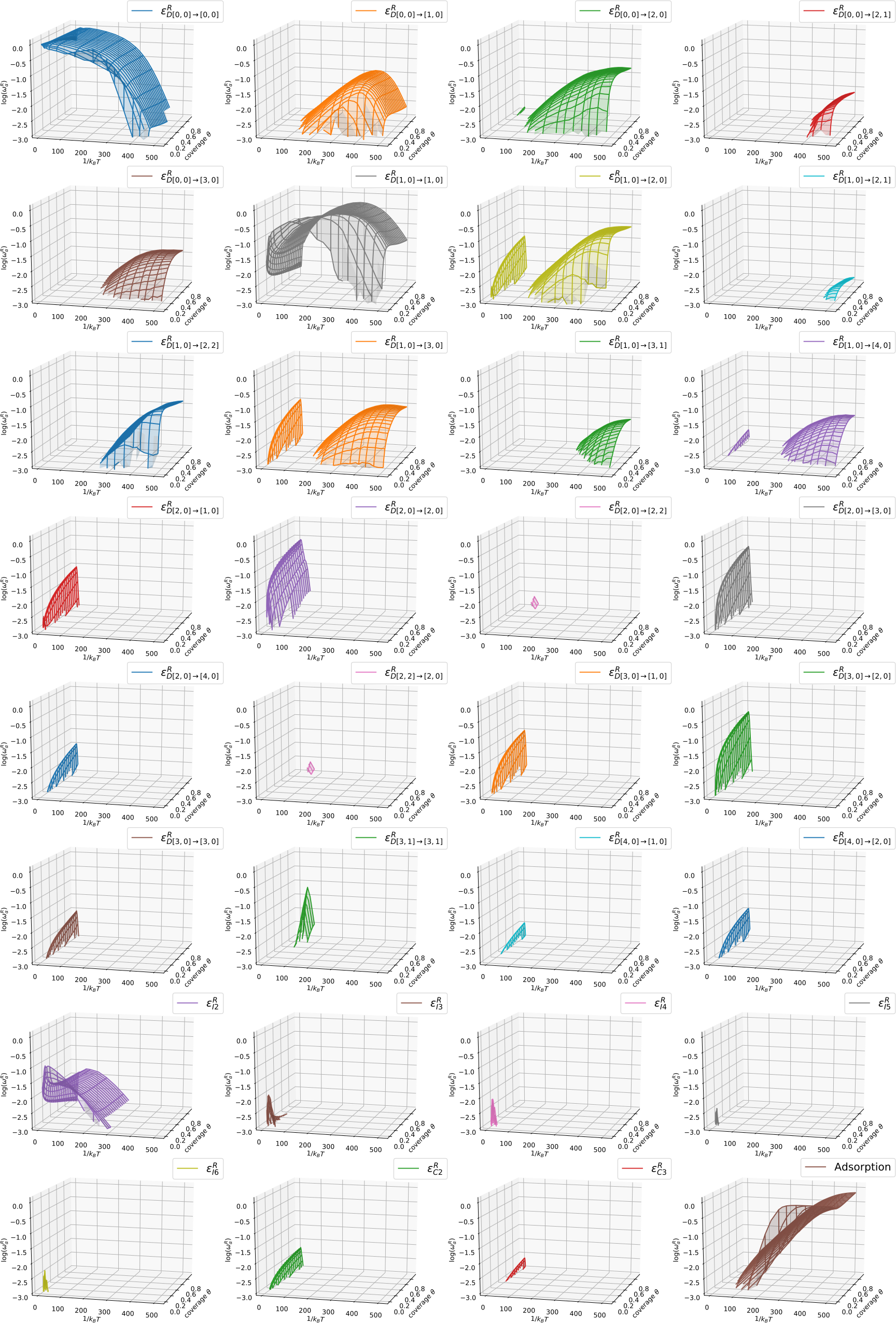}
  \caption{Event probabilities ($\omega_{\alpha}^{R}$) as a function
    of coverage and inverse temperature for the most relevant events
    in the Cu/Ni(111) system.}
  \label{fig:omegas3dcuni}
\end{figure}

\begin{figure}[htb!]
  \centering
  \includegraphics[width=0.85\columnwidth]{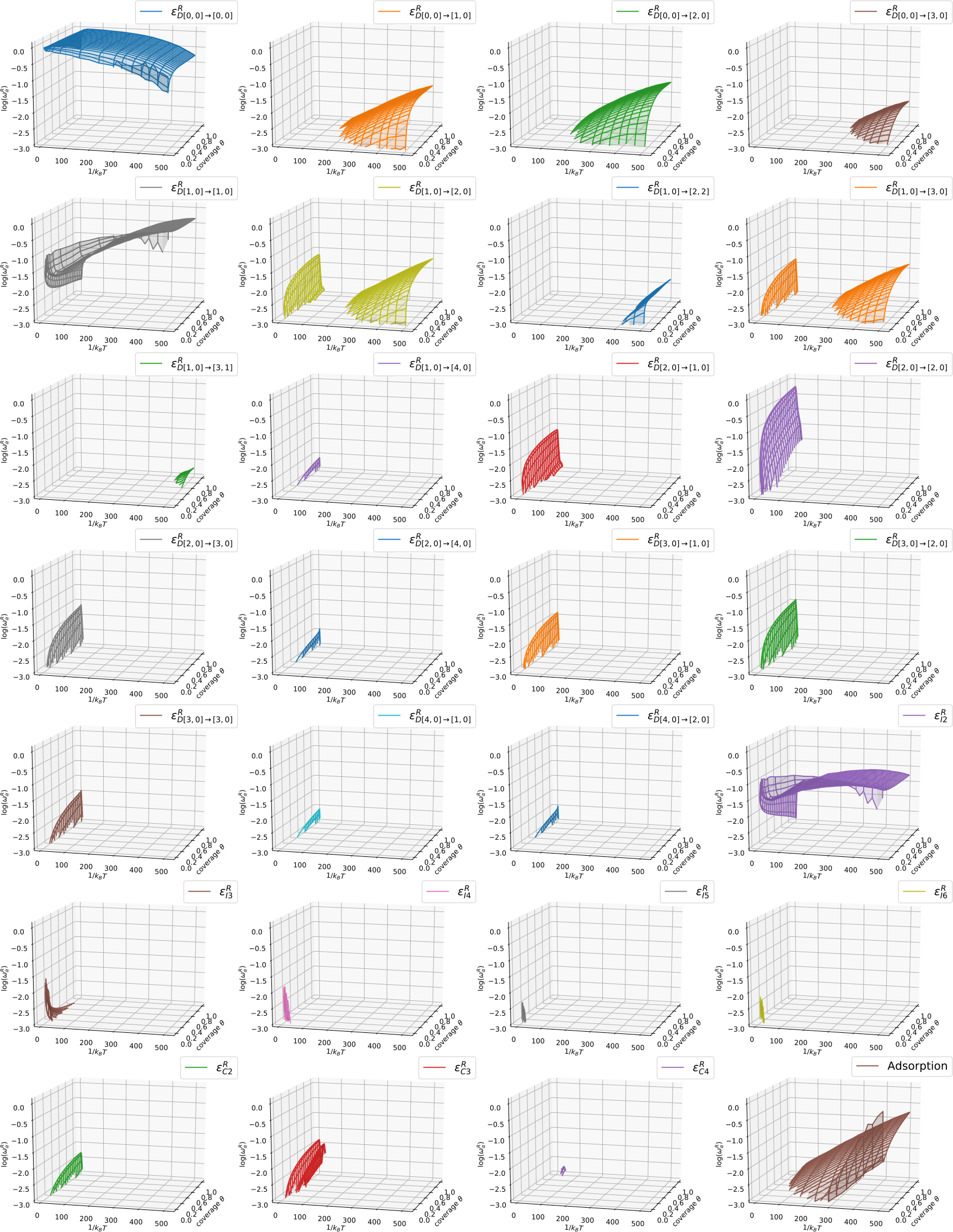}
  \caption{Event probabilities ($\omega_{\alpha}^{R}$) as a function
    of coverage and inverse temperature for the most relevant events
    in the Ni/Cu(111) system.}
  \label{fig:omegas3dnicu}
\end{figure}

\cleardoublepage
\newpage
\subsection{Input parameters}
Here we show the input parameters required to perform a typical
simulation of the present study.  The used Git revision of
``Morphokinetics'' is
\texttt{2b2811ae1187e69b3c55bf92ccb2c67c87761251}. The command to
compile the code is ``\texttt{ant jar}'' and the command to run a
simulation is ``\texttt{java -jar dist/morphokinetics.jar}''. The
latter must be executed within a folder containing the file
\texttt{parameters}, with the following content (removed the text
inside [ ]):
\begin{verbatim}
{
  "automaticCollections": true,
  "calculationMode": "concerted",
  "cartSizeX": 283,
  "cartSizeY": 283,
  "coverage": 100,
  "depositionFlux": 15000.0,
  "doIslandDiffusion": true,
  "doMultiAtomDiffusion": true,
  "forceNucleation": false,
  "justCentralFlake": false,
  "numberOfSimulations": 10,
  "outputData": true,
  "outputDataFormat": [
    {
      "type": "extra"
    },
    {
      "type": "ae"
    },
    {
      "type": "mko"
    }
  ],
  "psd": false,
  "randomSeed": false,
  "ratesLibrary": "CuNi", [or "NiCu"]
  "temperature": 23, [ranges from 23 to 1000]
  "withGui": false
}
\end{verbatim}

\end{document}